\documentclass[preprint]{aastex631}

\usepackage{amsmath}
\usepackage{multirow}

\begin{document}

\title{An Earth Encounter As the Cause of Chaotic Dynamics in Binary Asteroid (35107) 1991VH}

\correspondingauthor{Alex J Meyer}
\email{alex.meyer@colorado.edu}

\author[0000-0001-8437-1076]{Alex J. Meyer}
\affiliation{Smead Department of Aerospace Engineering Sciences, University of Colorado Boulder, 3775 Discover Dr, Boulder, CO 80303, USA}

\author[0000-0001-5875-1083]{Oscar Fuentes-Mu\~noz}
\affiliation{Smead Department of Aerospace Engineering Sciences, University of Colorado Boulder, 3775 Discover Dr, Boulder, CO 80303, USA}

\author{Ioannis Gkolias}
\affiliation{Aristotle University of Thessaloniki, Thessaloniki, Greece}

\author{Kleomenis Tsiganis}
\affiliation{Aristotle University of Thessaloniki, Thessaloniki, Greece}

\author[0000-0001-8434-9776]{Petr Pravec}  
\affiliation{Astronomical Institute of the Academy of Sciences of the Czech Republic, Fri\v{c}ova 298 \\
Ond\v{r}ejov, CZ 25165 Czech Republic}

\author{Shantanu Naidu}
\affiliation{Jet Propulsion Laboratory, California Institute of Technology, Pasadena, CA, USA}

\author[0000-0003-0558-3842]{Daniel J. Scheeres}
\affiliation{Smead Department of Aerospace Engineering Sciences, University of Colorado Boulder, 3775 Discover Dr, Boulder, CO 80303, USA}



\begin{abstract}

Among binary asteroids, (35107) 1991VH stands out as unique given the likely chaotic rotation within its secondary component. The source of this excited dynamical state is unknown. In this work we demonstrate that a past close encounter with Earth could have provided the necessary perturbation to allow the natural internal dynamics, characterized by spin-orbit coupling, to evolve the system into its current dynamical state. In this hypothesis, the secondary of 1991VH was previously in a classical 1:1 spin-orbit resonance with an orbit period likely between 28-35 hours before being perturbed by an Earth encounter within $\sim80,000$ km. We find if the energy dissipation within the secondary is relatively inefficient, this excited dynamical state could persist to today and produce the observed ground-based measurements. Coupled with the orbital history of 1991VH, we can then place a constraint on the tidal dissipation parameters of the secondary.

\end{abstract}



\section{Introduction} \label{sec:intro}

Near-Earth binary asteroids are typically comprised of an oblate primary rotating rapidly, with a tidally locked, elongated secondary in its orbit. These systems are generally thought to be formed by mass shedding from a rapidly rotating primary driven by YORP spin-up \citep{walsh2008rotational}, and tidal dissipation tends to relatively quickly synchronize the secondary's rotation with the orbit period \citep{taylor2014tidal}. In the synchronous state, these dissipative forces can eventually be balanced by radiative forces in the form of the binary YORP (BYORP) effect \citep{cuk2005effects}. This BYORP-tide singly-synchronous equilibrium is a steady state found among binary asteroids \citep{jacobson2011long}. The binary asteroid (35107) 1991VH is an exception to this singly-synchronous paradigm, as its secondary is not only in an asynchronous rotation state \citep{pravec2016binary}, it is most likely experiencing non-principal axis (NPA) rotation and undergoing an exchange of angular momentum between the secondary and the orbit \citep{naidu2015near,pravec2021photometric}. As a result, the orbit period and secondary rotation period have evolved over observations. This intriguing dynamical state has motivated substantial photometry and radar observations of the system. And with 1991VH's accessibility from Earth, it was originally selected as a target by NASA's SIMPLEx Janus mission \citep{scheeres2021janus}. As a result, there is a wealth of data and analysis surrounding this system. One question that remains is the source of the excited dynamical state of 1991VH, which we attempt to answer in this work.

Binary asteroids are characterized by strong spin-orbit coupling \citep{scheeres2006relative,scheeres2009stability}. These internal dynamics can easily excite the secondary into a complex rotational state given an external perturbation \citep{agrusa2021excited,meyer2023perturbed}. Fluctuations of the spin and orbit period of a system experiencing only principal axis rotation are generally not large enough to explain the variations seen in the observations \citep{meyer2021libration}. Thus, it seems the most natural explanation for the current dynamical state of 1991VH is NPA rotation within the secondary caused by an external perturbation and driven by the internal dynamics. This dynamical state has been postulated by \cite{pravec2016binary} and \cite{naidu2015near}, and in this work we provide a more quantitative analysis of this possibility.

The lifecycle of NEAs is chaotic and driven by orbital resonances with the giant planets and close encounters with the terrestrial planets \citep{liberato2023known}. Therefore a reasonable explanation for the excited dynamics of 1991VH is a recent close planetary encounter with the Earth. \cite{fuentes2022semi} showed that close Earth encounters with 1991VH are common, and it has also been shown that such close encounters can excite both the internal orbital and spin dynamics of binary asteroids \citep{fang2011binary,meyer2021effect}. Thus, it appears all the pieces are present to provide a satisfactory explanation of the current dynamical state of 1991VH. 

We test the hypothesis that 1991VH was previously in a singly-synchronous state, and possibly in the BYORP-tide equilibrium. This state is long-lived, but can be broken by resonances with the primary's spin \citep{wang2021break}, destabilizing impacts on the secondary \citep{cueva2024secular}, or planetary encounters \citep{merrill2024age}. These destabilizing events can lead to non-principal axis (NPA) rotation in the secondary, which has the effect of weakening or eliminating the BYORP effect \citep{quillen2022non,cuk2021barrel}. For the current asynchronous secondary rotation in 1991VH, the BYORP effect is not present.

In this work we specifically focus on the effects of a single close encounter with the Earth as a possible explanation for the unstable dynamics in 1991VH. This dynamical evolution has not been shown in the literature, and in the current work we are filling this gap. To achieve this, we perform a suite of Monte Carlo simulations, where a singly-synchronous 1991VH-like system is perturbed by the Earth in a variety of encounter geometries. We then compare the orbit period, secondary rotation period, semimajor axis, and eccentricity of the post-flyby system to check for similarities with the current observations.

We provide a background discussion of the observations and dynamics of 1991VH in Section \ref{sec:dynamics}. Then, we detail the results of the Monte Carlo flyby simulations in Section \ref{sec:encounters}. In Section \ref{sec:example} we perform a more detailed analysis of a single flyby case that produces a system very similar to 1991VH. We then give a detailed analysis of how long-lived the current dynamical state of 1991VH is expected to be in Section \ref{sec:dissipation}, and summarize and discuss implication in Section \ref{sec:discussion}.

\section{1991VH Observations and Dynamics} \label{sec:dynamics}

Observations of 1991VH date back to 1997 and have continued on through 2020, both through photometric lightcurves and radar range-Doppler imaging. This extensive set of observations has revealed the secondary in 1991VH to be in an asynchronous, likely chaotic spin state \citep{pravec2016binary,pravec2021photometric,naidu2015near}. Due to spin-orbit coupling, this spin state has resulted in an evolving orbit \citep{pravec2021photometric,meyer2022chaotic,meyer2022modeling}.

The observations of the secondary rotation period and orbit period are listed in Table \ref{tab:observations}. These measurements range between around 11.5 to 14.2 hours for the secondary spin period. Several estimates of the orbit eccentricity provide $e=0.05\pm0.02$ \citep{pravec2006photometric,naidu2012dynamics,naidu2018radar}. Estimates of the orbit semimajor axis range from $a=3.24$ km \citep{pravec2016binary} to $a=3.26$ km \citep{naidu2018radar}. 
Both photometric and radar observations give a volume-equivalent diameter of the primary of 1.2 km \citep{pravec2016binary,naidu2018radar}, although more recent observations suggest a smaller diameter of 0.9 km \citep{nugent2016neowise}. Radar observations reveal the primary to be top-shaped.

Photometric observations suggest the secondary has a diameter of 450 m with an elongation of $a_2/b_2=1.33\pm0.1$ \citep{pravec2016binary}, while radar observations give a more elongated value of $a_2/b_2=1.5$ \citep{naidu2015near}. Here, $a_2>b_2>c_2$ are the semiaxes of an ellipsoid fit to the shape of the secondary.

\begin{table}[ht!]
\caption{The observed secondary rotation and orbit period of 1991VH, calculated from lightcurves. Uncertainties are $\leq0.02$ hr. The bracketed value is not unique, and solutions exist between 7 and 29 hours. Orbit periods for all 2020 dates were determined from all the January-March data. Data taken from \cite{pravec2021photometric}.}
\label{tab:observations}
\centering
\begin{tabular}{ c c c }
 \hline
Epoch UT & Secondary Period (hr) & Orbit Period (hr) \\
 \hline
1997-03-17 & n/a & 32.69 \\
2003-02-25 & 12.84 & 32.63 \\
2008-07-01 & (14.18) & 32.8 \\
2020-01-27 & 11.57 & 32.50 \\
2020-02-18 & 11.78 & 32.50 \\
2020-02-26 & 11.55 & 32.50 \\
2020-03-02 & 11.61 & 32.50 \\
\hline
\end{tabular}
\end{table}

\subsection{Binary Asteroid Population} \label{subsec:population} 
The excited state of 1991VH is relatively unique among binary asteroids. Using data from \cite{pravec2016binary} which has been kept up-to-date in an online repository \citep{pravecdatabase}, we see only about $15\%$ of binary asteroids have asynchronous secondaries. This statistic is calculated using only systems which have a solution for both the orbit period and secondary rotation period.

\begin{figure}[ht!]
   \centering
   \includegraphics[width = 3in]{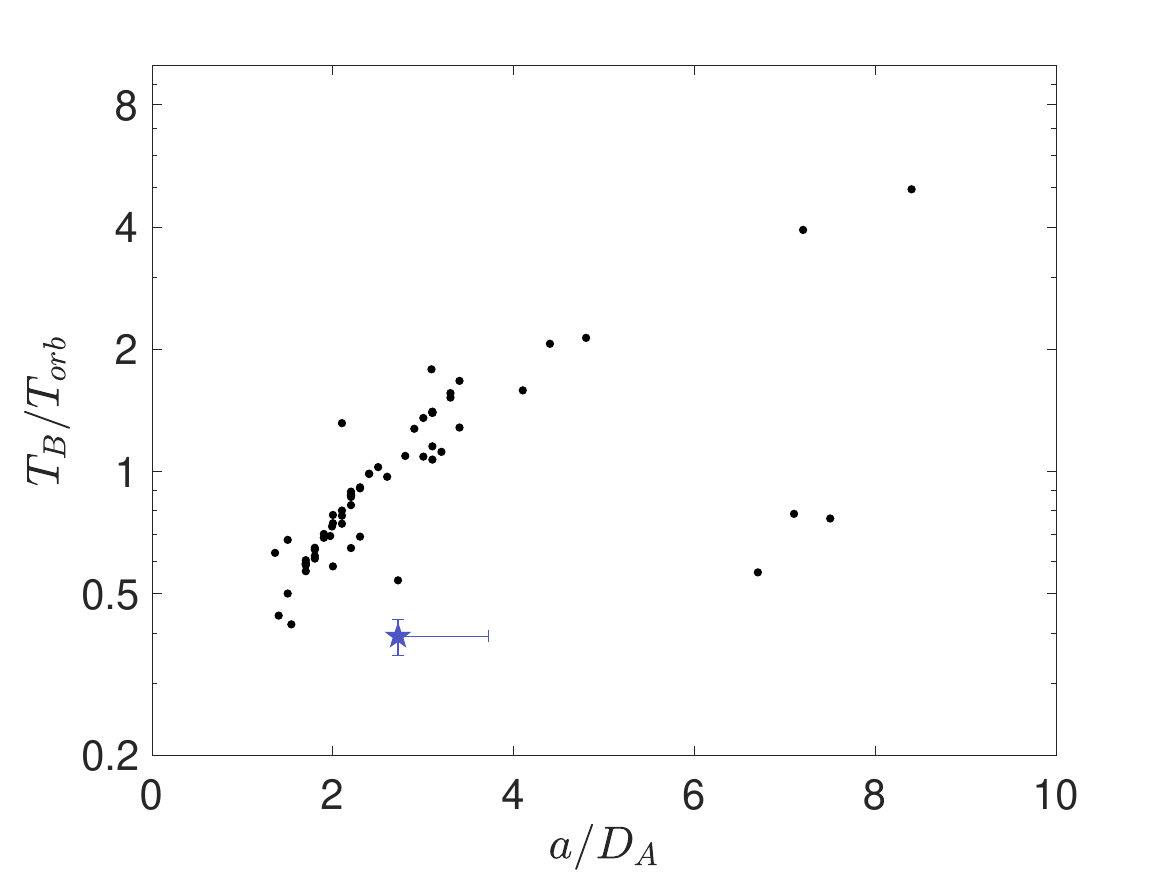} 
   \caption{The secondary spin to orbit period ratio as a function of the separation between the two asteroids. 1991VH is plotted as a star, and has a nominal separation $a/D_A\approx3$ and $T_B/T_{orb}\approx0.4$. Error bars show the range of possible values.}
   \label{fig:binary_pop}
\end{figure}

We demonstrate the uniqueness of 1991VH in Fig. \ref{fig:binary_pop}, in which we plot the secondary spin to orbit period ratio for the population of binary asteroids, as a function of the separation distance between the two asteroids. Among close binary asteroids, which we define here as having $a/D_A<5$, 1991VH has the smallest secondary to orbit period ratio (about 0.4). Here $a$ is the binary asteroid semimajor axis and $D_A$ is the diameter of the primary. 1991VH has a separation of about 3 primary diameters. Any of the systems which have a larger discrepancy between these periods are wide systems with orbit periods on the order of 100 hours. 

\subsection{Dynamical Model} \label{subsec:dynamics} 
To capture the spin-orbit coupling in binary asteroids one generally uses the full two-body problem (F2BP). This gives a 9 degree-of-freedom system, as the attitude of both bodies must be computed in addition to the relative position. We track the attitude of these bodies using the coordinate transformation matrices $\mathbf{C_A}$ and $\mathbf{C_B}$, which map vectors written in either the primary ($A$) or the secondary ($B$) body-fixed frame, respectively, to the inertial frame. In this environment, the mutual potential is
\begin{equation}
    U = G\int_A\int_B\frac{dM_AdM_B}{|\vec{r}-\mathbf{C_A}\vec{r}_A+\mathbf{C_B}\vec{r}_B|}
\end{equation}
where $\vec{r}$ is the relative position of body $B$ with respect to body $A$, $\vec{r}_A$ is the position of an infinitesimal mass element $dM_A$ of $A$ with respect to the body's barycenter, and similarly for $\vec{r}_B$ for $dM_B$ in $B$. The equations of motion for the F2BP are derived in \cite{maciejewski1995reduction}, and are implemented in the numerical tool General Use Binary Asteroid Simulator (\textsc{gubas}) \citep{davis2020doubly,davis2021gubas}. \textsc{gubas} integrates the F2BP using a user-specified expansion order for the mutual gravitational potential.

While \textsc{gubas} is an efficient implementation of the F2BP, we can simplify the problem further to allow for even faster calculation when appropriate. By reducing the primary to be spherical, we can ignore its attitude. This simplification means we will ignore the oblateness of the primary which can be important, but in this work we are more focused on the rotation state of the secondary. Using a second-order gravity expansion, this is equivalent to the sphere-ellipsoid problem \citep{scheeres2006relative}. In higher-fidelity simulations, we will remove this simplification and return to the full primary shape. But for now, we use the sphere-ellipsoid model, where the mutual potential energy can be approximated to second order using MacCullagh's formula \citep{murray2000solar}:
\begin{equation}
    U = -\frac{GM_AM_B}{r} - \frac{GM_A(I_x+I_y+I_z-3\Phi)}{2r^3}
\end{equation}
where
\begin{equation}
    \Phi = \frac{I_xx^2+I_yy^2+I_zz^2}{r^2}
\end{equation}
and $I_i$ is the principal inertia about axis $i$, and $I_x<I_y<I_z$. $x$, $y$, and $z$ define the location of the spherical primary in the body-fixed frame of the secondary.

The sphere-ellipsoid formulation allows for more efficient computation while still allowing for three-dimensional rotation of the secondary, which will be vital for understanding the chaotic dynamics of 1991VH. However, in a binary asteroid, the widely-separated assumption used in MacCullagh's formula may not be sufficiently accurate. Ideally, the separation would be at least an order of magnitude larger than the primary's radius, whereas for 1991VH the separation is only around 6-8 primary radii. For a more accurate calculation we turn to a more traditional spherical-harmonic approximation. For an ellipsoid, the spherical harmonics can be simply written up to degree and order 4 \citep{scheeres2016orbital},
\begin{align}
    C_{20} &= \frac{1}{5r_0^2}\left(c_2^2-\frac{a_2^2+b_2^2}{2}\right) \\
    C_{22} &= \frac{1}{20r_0^2}(a_2^2-b_2^2) \\
    C_{40} &= \frac{15}{7}(C_{20}^2+2C_{22}^2) \\
    C_{42} &= \frac{5}{7}C_{20}C_{22} \\
    C_{44} &= \frac{5}{28}C_{22}^2,
\end{align}
where $r_0$ is an arbitrary normalization radius. All other spherical harmonic terms are zero. These can be easily used in secondary-centered equations of motion using an algorithm such as the one defined in \cite{gottlieb1993fast}.

We will use MacCullagh's approach to obtain Poincaré maps for 1991VH in Sec. \ref{subsec:maps}, since the approximation is sufficient to qualitatively capture the behavior of the system and the computational efficiency allows us to generate more initial conditions for these maps. For the remainder of this work we will use the spherical harmonics approach.

Given the mass parameters of the system, the equilibrium spin rate in the sphere-ellipsoid model at a given separation distance $r$ is given by \citep{scheeres2009stability}:
\begin{equation}
    \dot{\theta}^2 = \frac{GM_AM_B}{r^3}\left(1+\frac{3}{2r^2}\left(\bar{I}_{y}+\bar{I}_{z}-2\bar{I}_{x}\right)\right)
    \label{eq:spin}
\end{equation}
where $\bar{I}_i$ is the mass-normalized inertia. This corresponds to the spin rate of the system when it is in the perfect 1:1 spin-orbit resonance.

\subsection{Dynamical Structure} \label{subsec:maps} 
We first explore the dynamical structure of 1991VH using a secondary with $a_2/b_2=1.3$, $b_2/c_2=1.2$, in the sphere-ellipsoid problem. This shape is the nominal estimate for the elongation of the secondary of 1991VH from \cite{pravec2016binary}. Since there is no information on the $b_2/c_2$ for 1991VH, we choose $b_2/c_2=1.2$ as this is near the values of other secondaries of near-Earth binary asteroids \citep{naidu2015radar,ostro2006radar}. We start the system in the synchronous equilibrium near its current dynamical state, then add eccentricity by perturbing the velocity such that the eccentricity becomes approximately 0.05.

\begin{figure}[ht!]
   \centering
   \includegraphics[width = 3in]{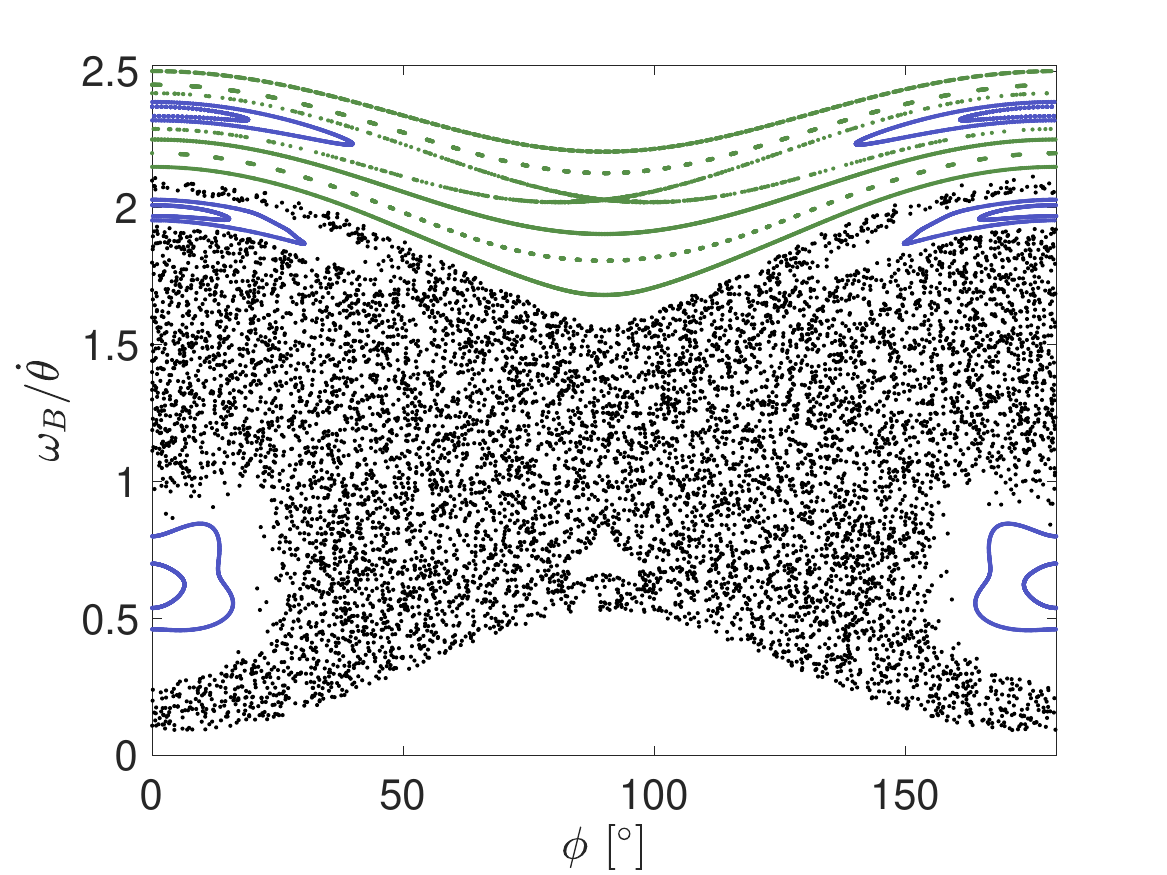} 
   \caption{A map of the sphere-ellipsoid problem limited to planar motion, plotting the normalized secondary spin rate as a function of the libration angle at periapsis. Periodic resonances are plotted in blue, quasi-periodic curves in green, and chaotic motion in black.}
   \label{fig:poincare1}
\end{figure}

We test different values of the secondary spin rate, keeping the velocity perturbation constant so the translational kinetic energy is the same for each initial condition. Following \cite{wisdom1984chaotic}, we plot maps of the normalized secondary spin rate $\omega_B/\dot{\theta}$ as a function of the secondary's libration angle $\phi$ at each periapsis crossing. In binary asteroid dynamics, the periapsis is not necessarily well defined, as the Keplerian orbit precesses at a rate comparable to the orbit rate \citep{scheeres2009stability} In these plots, we define a periapsis crossing as a local minimum in the orbit separation. Thus, we adopt a purely numerical definition for the periapsis crossing. We define the libration angle as the angle between the secondary's longest axis and the vector from the secondary to the primary. The results, restricted to planar motion, are shown in Fig. \ref{fig:poincare1}.

These Poincare maps show spin-orbit resonances in blue, quasiperiodic dynamics in green (in which the Poincare map will trace out a curve but never exactly repeat), and chaotic dynamics in black. This demonstrates how different initial conditions result in varying dynamical behavior.

These results are similar to the results obtained by \cite{naidu2015near}, which is expected as our approach is nearly the same. We see stable equilibria around the 1:1, 2:1, and 5:2 spin-orbit resonances. However, the perturbation is strong enough to force overlapping of the 2:1 and 3:2 resonances, leading to a very large chaotic sea. Several distinct initial conditions result in the chaotic sea shown in Fig. \ref{fig:poincare1}. Above this chaotic sea, the system follows predictable quasi-periodic curves. We also note the resonances are shifted away from exactly 1, 2, or 2.5 due to the effect of libration in the system, which shifts the secondary's spin rate at periapsis \citep{naidu2015near}.

Around the 1:1 spin-orbit resonance of the full 3 dimensional problem, \cite{agrusa2021excited} showed that resonances among the natural frequencies of the system can excite out-of-plane rotation. Thus, certain shapes of the secondary are inherently unstable. The large uncertainty on the secondary's shape in 1991VH allows for instabilities in the system driven by the internal spin-orbit coupling in binary asteroids. Notably, the nominal shape of $a_2/b_2=1.3$, $b_2/c_2=1.2$ is near a 2:1 resonance between the nutation and libration of the secondary, and as we will show next, this system is susceptible to chaotic motion.

\begin{figure}[ht!]
   \centering
   \includegraphics[width = 3in]{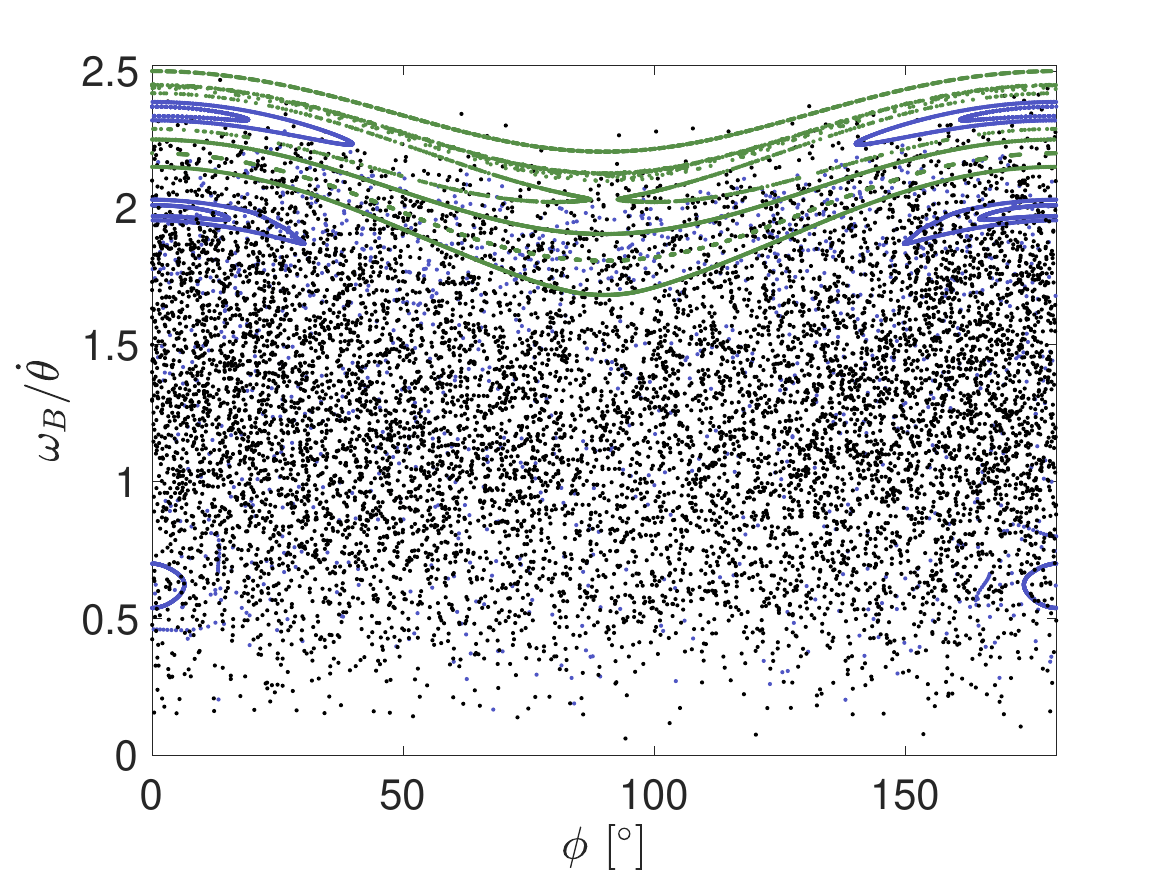} 
   \caption{A map of the sphere-ellipsoid problem perturbed into three-dimensional rotation, plotting the normalized secondary spin rate as a function of the libration angle at periapsis. The colors match the same initial conditions as in Fig. \ref{fig:poincare1}.}
   \label{fig:poincare2}
\end{figure}

While Fig. \ref{fig:poincare1} presents a classical result in the spin-orbit problem, it is limited to planar motion. However, it is well known that out-of-plane rotation in the secondary is important in binary asteroids. Furthermore, observations suggest the secondary of 1991VH could be in a state of non-principal axis (NPA) rotation. So we expand our analysis to include out-of-plane rotation. Taking the same initial conditions as those in Fig. \ref{fig:poincare1}, we add now a small perturbation to the secondary's spin axis of $(1\times10^{-6})^\circ$ along its minimum principal inertia direction. The results of these simulations are shown in Fig. \ref{fig:poincare2}.

We see a significant change to the structure of the dynamics. Most importantly, because we have increased the dimensionality by 2, the resonant and quasi-periodic curves of the two dimensional problem are no longer barriers to chaotic motion in three dimensions. We see some solutions around the 1:1 resonance become chaotic, even though the spin perturbation is very small. The chaotic sea now crosses the 1:1, 2:1, and 5:2 resonances, along with the quasiperiodic curves. However, these quasiperiodic curves and the 2:1 and 5:2 resonances generally keep the same structure as in the planar case. These solutions have sufficient angular momentum that the small spin perturbation is not large enough to cause these solutions to become unstable. The instabilities we see in Fig. \ref{fig:poincare2} are not unexpected, as a linear stability analysis predicts this behavior for a wide range of secondary shapes \citep{wisdom1984chaotic}.

This analysis demonstrates the importance of considering full three-dimensional rotation in dynamical analyses of 1991VH, and binary asteroids in general. Even very small perturbations can destroy the structure that exists when the system is restricted to planar motion and rotation.

Applied specifically to 1991VH, this shows that a very small perturbation to the orbit, sufficient to produce the observed eccentricity, could kick-start a period of NPA rotation, even if the system was previously in the 1:1 spin-orbit resonance. This provides one possible evolutionary pathway to place 1991VH in the dynamical state observed today.

\subsection{External Perturbation} \label{subsec:perturbations} 
The most likely source of external perturbation to 1991VH is a close gravitational encounter with a terrestrial planet. Over the last 100,000 years, there is around a 25\% chance that 1991VH had a close encounter with either Mars or Earth sufficient to increase the eccentricity to 0.1. \citep{fuentes2022semi}.

On average, the change in eccentricity to a binary asteroid can be calculated by \citep{fang2011binary}:
\begin{equation}
    |\Delta e| \approx 1.89\sqrt{\frac{G}{M_A+M_B}}\frac{M_Sa^{3/2}}{v_\infty q^2}
\end{equation}
where $M_S$ is the mass of the perturbing planet, $a$ is the semimajor axis of the binary, $v_\infty$ is the hyperbolic excess speed of the flyby, and $q$ is the closest approach distance between the planet and the binary's barycenter. Using a semimajor axis of 3.25 km and a hyperbolic excess velocity of 10 km/s, 1991VH would gain an eccentricity of 0.05 from a close Earth encounter with a periapsis of approximately $160,000$ km. From the semi-analytical propagation carried out by \cite{fuentes2022semi}, such an encounter was possible no sooner than $12,000$ years ago.

The average change in semimajor axis is similarly written as \citep{fang2011binary}:
\begin{equation}
    |\Delta a| \approx 1.48\sqrt{\frac{G}{M_A+M_B}}\frac{M_Sa^{5/2}}{v_\infty q^2}.
\end{equation}
Using the same numbers for a $160,000$ km Earth flyby, this corresponds to a change in the semimajor axis of about 140 m. Thus, on average, we would expect the pre-encounter 1991VH to have a semimajor axis around 3.22-3.28 km, with the flyby either increasing or decreasing the semimajor axis to its currently estimated value.

In their high-fidelity simulations of a 1991VH-like binary asteroid undergoing an Earth encounter, \cite{meyer2021effect} find a flyby distance of $100,000$ km results in an eccentricity of around $0.05\pm0.01$. This also corresponds to a change in the binary's mutual semimajor axis of about $100\pm100$ m. The binary asteroid parameters used in that work are slightly different from those of 1991VH, but the results are still generally consistent with the analytic predictions.

\section{Earth Encounter Simulations} \label{sec:encounters}

Our hypothesis is that 1991VH was previously in a singly-synchronous state, similar to the majority of observed binary asteroids \citep{pravec2016binary}, before a close Earth encounter added eccentricity to the system and excited the system into its currently-observed chaotic state. To test this hypothesis, we vary possible pre-encounter orbit periods for the singly-synchronous state, then simulate a variety of Earth encounter conditions. 

\subsection{Simulation Inputs} \label{subsec:inputs} 
In these simulations, we fix the bulk density of the system to be $\rho=1.6 $g/cm$^3$ \citep{naidu2015near}, and set the primary and secondary densities equal to one another. Thus, when we choose the pre-encounter orbit period, this fixes the pre-encounter semimajor axis. Instead of using the Keplerian relationship to determine the semimajor axis, we adopt the approach used by \cite{meyer2023perturbed}, who demonstrated the inaccuracy of the Keplerian elements in binary asteroid dynamics due to relaxing the point-mass assumption. Thus, we use the stroboscopic orbit period, also called the plane-crossing orbit period, which is determined by the differences in timings of the secondary crossing of an arbitrary inertial plane. Equivalently, this is the time required for the secondary to complete one full revolution around the primary in inertial space. Following \cite{meyer2023perturbed}, we iteratively calculate the separation distance required to achieve a desired stroboscopic orbit period using a numerical secant search algorithm. 

In these simulations, we will use the F2BP dynamics, and thus allow for the primary to be aspherical. We set the primary $J_2=0.02$ (S. Naidu, personal communication, April 2024) and assume it is axisymmetric, and use a second degree-and-order expansion of the mutual potential. We fix the primary volume-equivalent diameter to 1.2 km. 

\begin{figure}[ht!]
   \centering
   \includegraphics[width = \textwidth]{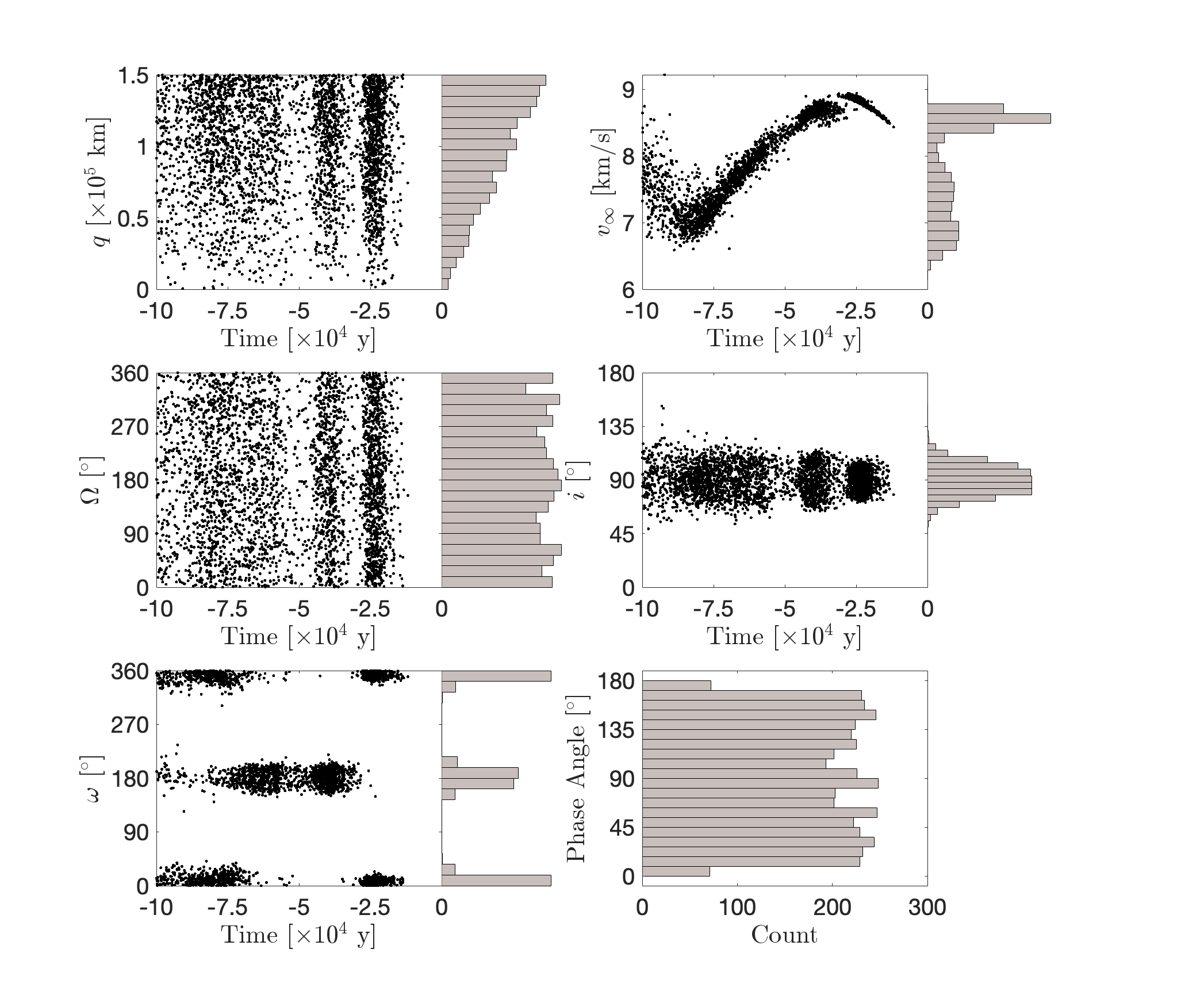} 
   \caption{The close approach distance and hyperbolic excess speed, along with the Keplerian angles of the hyperbolic flyby trajectories as a function of time in the past. We also plot their overall distributions as histograms. These flybys are calculated by propagating the heliocentric orbit of 1991VH into the past and finding close approaches with Earth. We also plot the distribution of Earth-secondary phase angle at closest approach used in our Monte Carlo analysis.}
   \label{fig:flyby_elements}
\end{figure}

We generate realistic flyby geometries of 1991VH relative to the Earth by propagating its orbit backward in time, taking into account the uncertainties in its ephemeris. We do this using the semi-analytical NEO propagation tool developed by \cite{fuentes2022semi}. We use the heliocentric orbit nominal solution and full covariance from JPL's SSD/CNEOS Small-Body Database\footnote{Orbit solution retrieved from JPL's SSD/CNEOS SBDB API available at \url{https://ssd-api.jpl.nasa.gov/doc/sbdb.html}. Data accessed 03-08-2024.}. This solution uses a 29 year data arc, including 91 high-precision observations from the Gaia Focused Product Release \citep{david2023gaia}, incorporated in the solution following the data treatment of \citet{fuentes2024fpr}. Even though the latter observations further constrain 1991VH's orbit, the long-term propagation leads to a statistical distribution of the Earth flyby parameters. The results of this propagation for the past $100,000$ years are shown in Fig. \ref{fig:flyby_elements}, showing every close approach within $150,000$ km of Earth. In the propagation we occasionally find close encounters with Mars and Venus, but based in the estimated perturbation \citep{fang2011binary} and the lower frequency of those encounters \citep{fuentes2022semi} we exclude them from this analysis.

There are a cluster of possible flybys around $25,000$ years ago, followed by another cluster around $40,000$ years ago. After these periods, uncertainties grow large enough that we have a more uniform distribution in flybys. The distribution of the longitude of the ascending note is generally uniform for all flybys, but the hyperbolic excess velocity has a clear evolution with time, generally decreasing into the past until around $80,000$ years ago. The inclination of the flybys appears roughly Gaussian, and centered around $90^\circ$. The argument of periapsis is centered around either $0/360^\circ$ or $180^\circ$. We generate a uniform distribution of the Earth-secondary phase angles at closest approach, indicating we are testing a wide range of possible geometries Since there is no data on the system's phase angle, we randomly draw from all possible values.

For the pre-encounter binary asteroid, we use the range of possible values outline in Table \ref{tab:inputs}. This gives the pre-encounter orbit period, as well as the size and shape of the secondary. The separation distance is calculated from the orbit period such that the secondary is in a physically circular orbit, using the F2BP dynamics to eliminate pre-encounter libration from the secondary.

\begin{table}
\caption{The physical parameters used in the Monte Carlo simulations of Earth close encounters.}
\label{tab:inputs}
\centering
\begin{tabular}{ c c c }
 \hline
Parameter & Symbol & Range \\
 \hline
Pre-encounter Binary Orbit Period & $T_{pre}$ & 27-37 hr \\
Secondary Diameter & $D_{B}$ & 400-500 m \\
Secondary Prolateness & $a_2/b_2$ & 1.1-1.5 \\
Secondary Oblateness & $b_2/c_2$ & 1.1-1.5 \\
\hline
\end{tabular}
\end{table}

We use the Spherical-Restricted Full 3-Body dynamical model defined and implemented by \cite{meyer2021effect}. This model calculates the full spin-orbit coupling between the primary and secondary while also influenced by a distant and large spherical perturber, in our case the Earth. We simulate the flyby for 20 days with closest approach centered at 10 days. After this, we hand off the results to the pure F2BP to speed up computational time without needing to track the effect of the Earth, and simulate the system for a full year. \cite{meyer2021effect} showed that plus or minus 2 days from closest approach, the effect of the perturbing planet was negligible, so we are conservative in our use of plus and minus 10 days from closest approach.

\subsection{Results} \label{subsec:results} 
As already discussed, \cite{meyer2023perturbed} showed the inaccuracy of Keplerian elements in representing binary asteroid orbits. As such, we will also adopt the use of the `observable' semimajor axis and eccentricity in our results. These quantities are defined using only the maximum and minimum separation distances within some time interval; in this work we will use a sliding time window of 5 days.

This still leaves the question of how best to calculate the secondary's spin period, which is a key piece of information provided by observations. We are concerned with the dynamics of 1991VH that will reproduce these observations, so our approach should mimic how the spin period is calculated in these observations. Generally this is done using lightcurves of the secondary \citep{pravec2016binary}. We define synthetic lightcurves, which we will calculate using the visible cross section of the secondary's cross-sectional area:
\begin{equation}
    A = \pi\sqrt{\hat{x}^2b_2^2c_2^2+\hat{y}^2a_2^2c_2^2+\hat{z}^2a_2^2b_2^2}
\end{equation}
where $\hat{x}$, $\hat{y}$, and $\hat{z}$ is the direction of the observer defined in the body-fixed coordinates of the secondary. 

After calculating the cross-section area, we perform a 1-degree Fourier fit over the same 5-day sliding observation window. If the r-squared value of the fit is at least 0.5, we keep the fit as a measurement of half the secondary spin period. Due to the symmetry of the perfect ellipsoid, the Fourier fit will only measure half the spin period, so we correct by a factor of 2 to obtain the true spin period.

Our observer is fixed at an arbitrary position in inertial space, so these synthetic lightcurves calculate the sidereal period. The real period measurements are reported as synodic values, so we do not have an exact comparison between the synthetic and true data. However, the difference between the sidereal and synodic periods are less than 0.1 hour for all observations \citep{pravec2006photometric,pravec2021photometric}, so this difference is small. Due to the uncertainties and chaotic dynamics in the system, we will not be able to exactly reproduce the true measurements, so this difference is acceptable.

\begin{figure}[ht!]
   \centering
   \includegraphics[width = 3in]{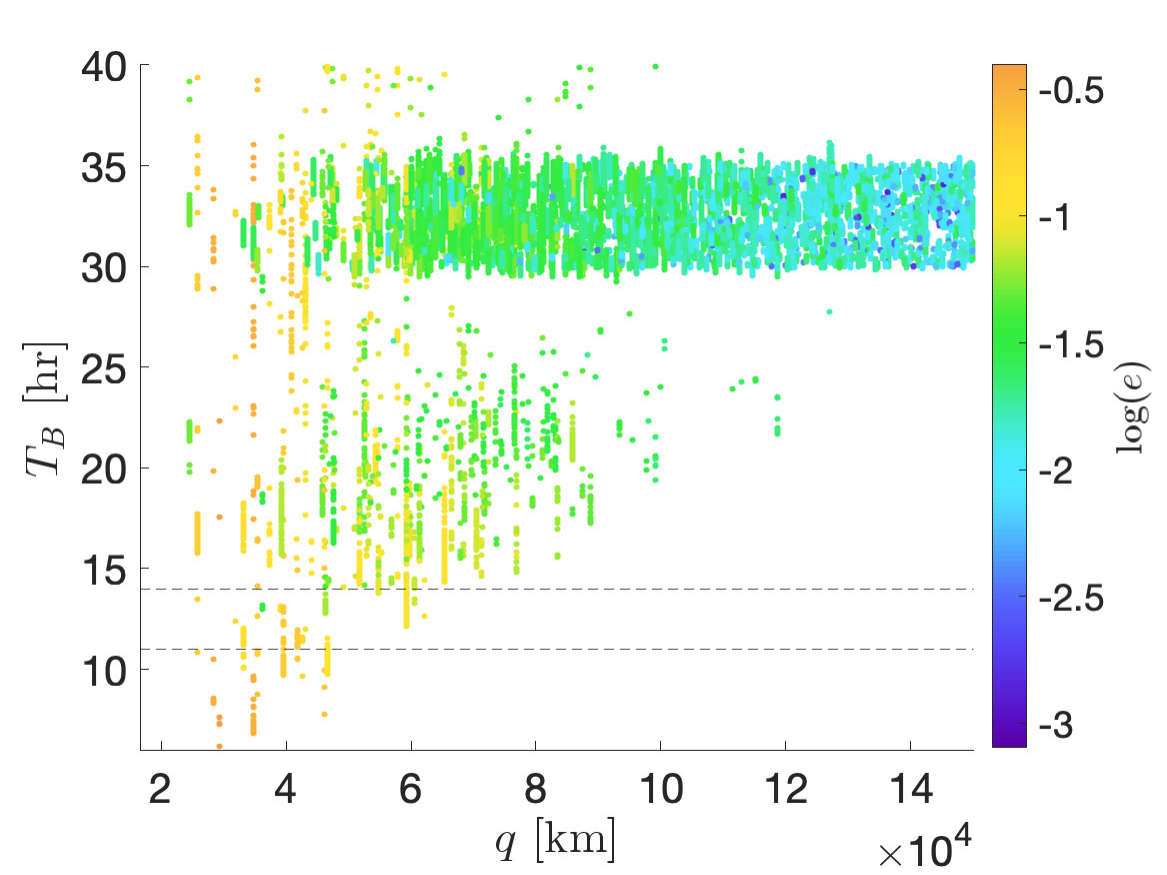} 
   \caption{The secondary spin periods as a function of Earth closest approach. The dashed lines bracket the range of observed values, and the color gradient shows eccentricity. The observed eccentricity of 1991VH is between 0.03 and 0.07.}
   \label{fig:period_rp_scatter}
\end{figure}

We perform $4,600$ Monte Carlo simulations of Earth close encounters, randomly sampling from the distribution in Fig. \ref{fig:flyby_elements}. In each of these simulations we calculate the stroboscopic orbit period of the post-encounter system. To match the observations, we remove any results that have an average orbit period less than 30 hours or greater than 35 hours. After this removal, we obtain $1,866$ results, which are plotted in Fig. \ref{fig:period_rp_scatter}. This plots the secondary spin period measurements as a function of closest approach distance, with dashed lines bracketing the observed values.

From Fig. \ref{fig:period_rp_scatter}, we see a large grouping of secondary spin periods in the 30-35 hour range. These are the systems which remain synchronous with the orbit period after the flyby. We see most simulations fall in this category, but a significant number of simulations do have faster or shorter secondary spin periods, indicating these systems have become asynchronous. The number of results falling within the observed window, bracketed by dashed lines, is small but significant, indicating it is possible to reproduce the observed systems after a single Earth close encounter.

We see a narrow range of flyby conditions that are able to reproduce both the observed secondary rotation period and the orbital eccentricity. These flybys have a close approach distance within the range of $50,000$ - $80,000$ km. Closer than this produces an eccentricity that is too large, while further than this does not provide a sufficient perturbation to induce asynchronous rotation in the secondary. This range is smaller than the analytic prediction in Section \ref{subsec:perturbations}. This suggests reproducing the observed secondary spin period is more difficult than simply achieving the observed eccentricity.

\begin{figure}[ht!]
   \centering
   \includegraphics[width = 3in]{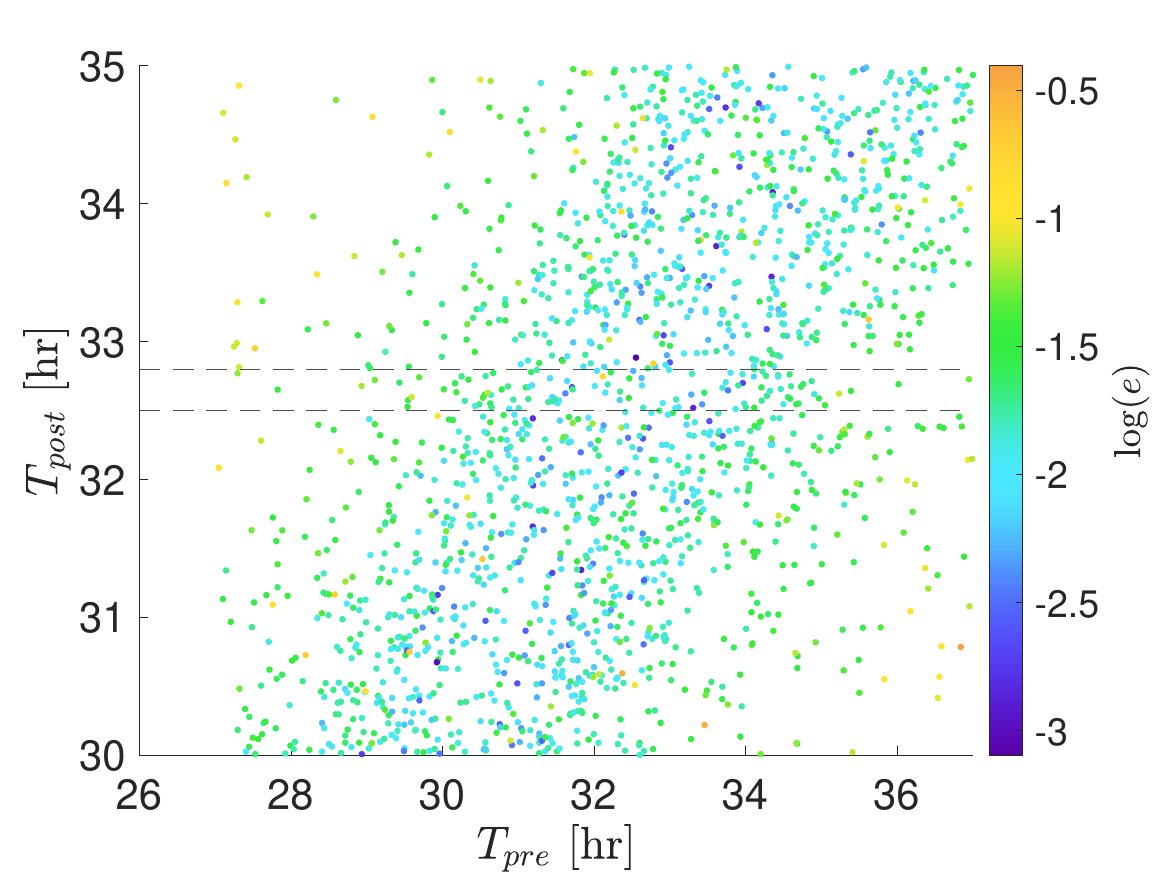} 
   \caption{The post-encoutner orbit period as a function of pre-encounter orbit period. The dashed lines bracket the range of observed values, and the color gradient shows eccentricity. The observed eccentricity of 1991VH is between 0.03 and 0.07.}
   \label{fig:period_scatter}
\end{figure}

In Fig. \ref{fig:period_scatter}, we plot the post-encounter orbit period as a function of pre-encounter orbit period. Again, the dashed lines bracket the observed values. In order to achieve these observed values, the pre-encounter orbit period of the system was likely between 28 and 35 hours, but to match the observed eccentricity, the pre-encounter orbit period was likely not within the 30-33 hour range. Thus, the most likely pre-encounter orbit period was either 28-30 or 33-35 hours. This indicates the orbit period and semimajor axis of 1991VH have likely not changed significantly as a result of a possible Earth encounter in its past, and the main change is the eccentricity and rotation of the secondary.

In our simulations, we do not find a single case in which a system remains in planar rotation and has a secondary rotation period near the observed values. This is another indication of chaotic rotation in 1991VH.

\section{Example History} \label{sec:example}

Here we present an example simulation providing a close match with observations. This system was initially in a circular orbit with a separation distance of 3.11 km and an orbit period of 29.9 hours. The secondary ellipsoid has a volume-equivalent diameter of 458 m, and axis ratios $a_2/b_2=1.39$ and $b_2/c_2=1.17$. This system undergoes an Earth encounter with a close approach distance of $46,580$ km with a hyperbolic excess speed of 7.6 km/s. The hyperbolic elements are $i=78^\circ$, $\Omega=217^\circ$, and $\omega=193^\circ$. 

\begin{figure}[ht!]
   \centering
   \includegraphics[width = \textwidth]{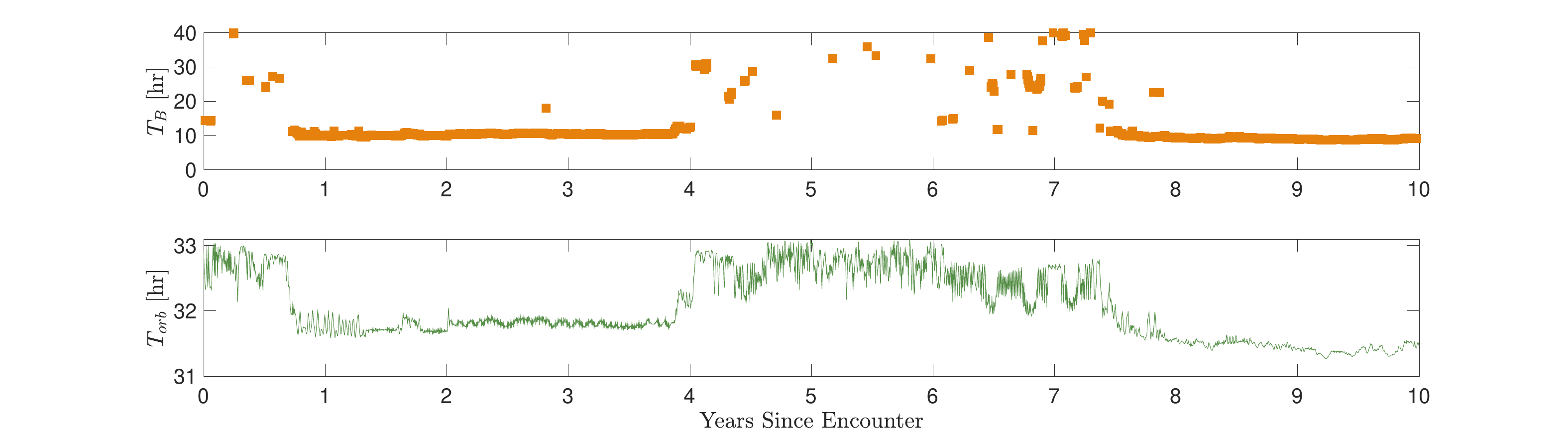} 
   \caption{The fitted secondary spin periods (top) and stroboscopic orbit period (bottom) for one example simulation over 10 years after an Earth encounter.}
   \label{fig:example_period}
\end{figure}

To study this history, we integrate the system forward for 10 years after the flyby and plot the post-encounter secondary spin period and orbit period in Fig. \ref{fig:example_period}. The results are not a perfect match, but lie very close to the observed values. The secondary spin period alternates between epochs of more chaotic rotation and quasi-constant spin. The periods of quasi-constant spin last for several years at around a 10 hour spin period, generally consistent with observations. During the more chaotic epochs the spin period varies between 10 and 40 hours, and larger values are also possible. We note the actual observations of the secondary's rotation period in 1991VH do not see these longer rotation periods, although there is a possible estimate of a spin period up to 29 hours (see Table \ref{tab:observations}).

Likewise, the orbit period fluctuates between around 31.5 and 33 hours. The observed orbit period ranges between 32.5 and 32.8 hours. Thus, our simulated orbit period is not an exact match, but very close. There is a clear correlation with the secondary spin period. During the quasi-constant secondary spin period, the orbit period is slightly less than 32 hours, but increases during the periods of chaotic rotation.

\begin{figure}[ht!]
   \centering
   \includegraphics[width = \textwidth]{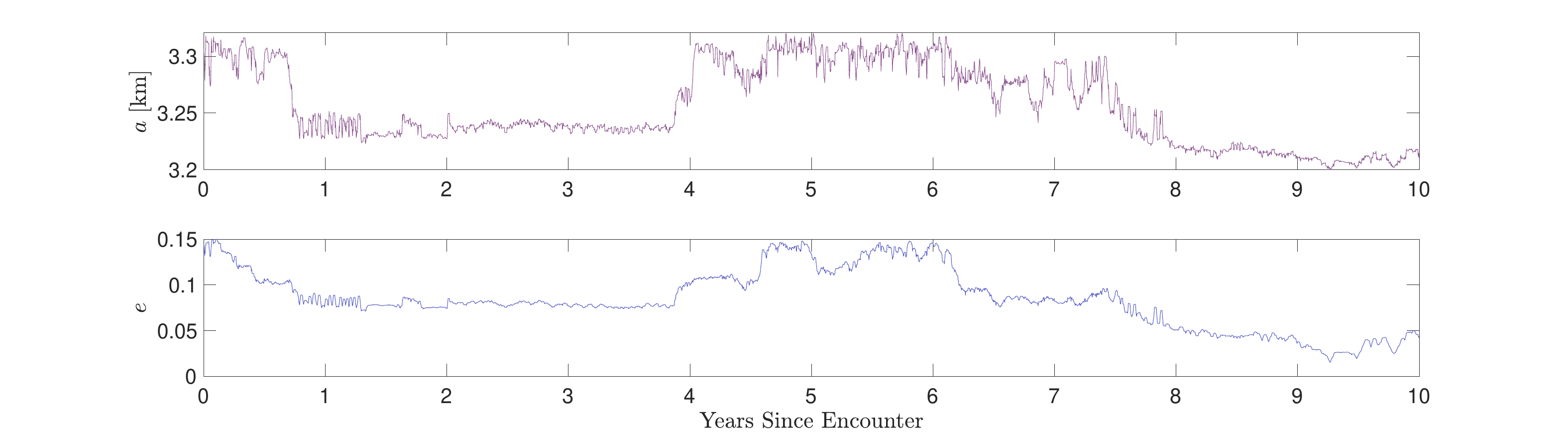} 
   \caption{The observable semimajor axis (top) and eccentricity (bottom) for one example simulation over 10 years after an Earth encounter.}
   \label{fig:example_orbit}
\end{figure}

In Fig. \ref{fig:example_orbit} we plot the observable semimajor axis and eccentricity. The semimajor axis ranges between 3.23 and 3.32 km. Similar to the orbit period, these values are a close match to the true observations of 32.4-32.6 km. Again there is a correlation with the secondary spin, with a lower semimajor axis during quasi-constant spin and a larger value during the chaotic rotation.

The eccentricity ranges between 0.02 and 0.15. The true observed eccentricity values range from 0.03 to 0.07. The periods of higher eccentricity correlate the the chaotic rotation and lower eccentricity during quasi-constant spin. So while our simulated eccentricity values can exceed these observations, they are overall generally consistent with observations.

\begin{figure}[ht!]
   \centering
   \includegraphics[width = \textwidth]{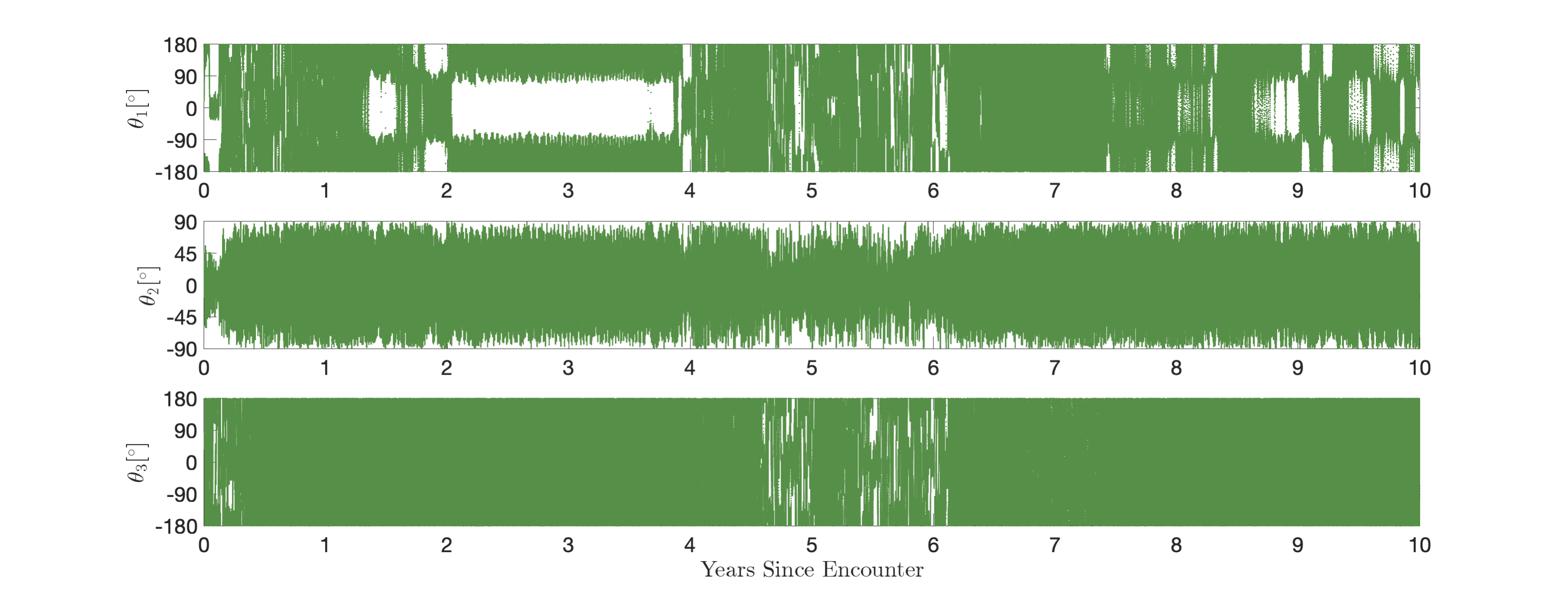} 
   \caption{The roll (top), pitch (middle), and yaw (bottom) angles tracking the secondary's orientation relative to the singly-synchronous equilibrium.}
   \label{fig:example_euler}
\end{figure}

We track the secondary's orientation with a classical set of 1-2-3 roll, pitch, yaw Euler angles relative to the rotating Hill frame with $z$-axis aligned with the orbit's angular momentum vector. In this construction, a tidally-locked secondary would have zero rotation in all three axes. Fig. \ref{fig:example_euler} shows these Euler angles over time for this example. During the period of quasi-constant spin, the secondary is generally only librating in its roll angle, either around $0^\circ$ or $180^\circ$. However, the other angles are circulating, indicating even in this period of quasi-constant spin, the secondary is still fully tumbling. During the periods of chaotic rotation, all three angles are circulating. 

In this simulation, the secondary enters a state of fully chaotic rotation, and in particular the barrel instability is prominent in the roll angle \citep{cuk2021barrel}. As this dynamical state closely reproduces the observations, this is another indicator that 1991VH is indeed experiencing chaotic rotation in the secondary. 

Overall, this simulation demonstrates that due to the chaotic nature of binary asteroids, it is possible for a single Earth encounter to transform an equilibrated, singly-synchronous binary asteroid into a system that looks similar to the current 1991VH.

\section{Secular Energy Dissipation} \label{sec:dissipation}

While we have demonstrated a close Earth encounter can create a system similar to 1991VH from a previously singly-synchronous binary asteroid, it is unclear how long-lived this excited state will last. The earliest Earth encounter sufficiently close to produce these dynamics was around 12,000 years ago \citep{fuentes2022semi}.

In the singly-synchronous configuration, for a given spin rate of the primary the system's energy is minimized for the amount of angular momentum in the system. A perturbation such as a close planetary flyby will change the energy and angular momentum such that NPA rotation is allowable, precipitating an exchange of angular momentum between the secondary and the orbit. Since the secondary's angular momentum is much smaller than that of the orbit and primary, a small excess in energy at a given angular momentum level allows for NPA rotation.

In the literature, it has been argued that NPA rotation can be long-lived \citep{wisdom1984chaotic}, including specifically for binary asteroids \citep{quillen2022non}. In their analysis, \cite{naidu2015near} make arguments that the asynchronous state of 1991VH can also be persistent over long times. However, recently \cite{meyer2023energy} demonstrated that eccentricity damping in binary asteroids can be much faster than classical analytic predictions due to the spin-orbit coupling and the relationship between libration and eccentricity. In this section, we provide a more concrete analysis to investigate the question of the persistence of the NPA rotational state of 1991VH.

In the minimum-energy configuration, the equilibrium spin rate is defined by Eq. \ref{eq:spin}. Because the secondary is in the 1:1 spin-orbit resonance in this configuration, its spin rate relative to an inertial frame is also defined by \ref{eq:spin}. Thus, the angular momentum for this configuration is simply written as:
\begin{equation}
    H=(I_{Bz}+Mr^2)\sqrt{\frac{G(M_A+M_B)}{r^3}\left(1+\frac{3}{2r^2}\mathcal{I}\right)}+I_{Az}\omega_A
    \label{eq:H}
\end{equation}
where
\begin{equation}
    \mathcal{I}=\bar{I}_{By}+\bar{I}_{Bz}-2\bar{I}_{Bx}
\end{equation}
and 
\begin{equation}
    M=\frac{M_AM_B}{M_A+M_B}.
\end{equation}
Here we are explicitly assuming the primary is limited to principal-axis rotation about its major axis.

In the equilibrium configuration, the orbital velocity is $v=r\dot{\theta}$. Using this, the corresponding minimum energy can be written as \citep{scheeres2009stability}:
\begin{equation}
    E^* = \frac{1}{2}(I_{Bz}+Mr^2)\frac{G(M_A+M_B)}{r^3}\left(1+\frac{1}{3r^2}\mathcal{I}\right)-\frac{GM_AM_B}{r}\left(1+\frac{1}{2r^2}\mathcal{I}\right)+\frac{1}{2}I_{Az}\omega_A^2.
    \label{eq:E}
\end{equation}

When the system is perturbed away from this equilibrium, it will dissipate energy at a constant angular momentum. As the orbit expands and the primary's rotation rate slows, it will eventually be constrained to synchronous rotation in the secondary. This happens at the combination of separation and primary spin rate defined in Eq. \ref{eq:E}. At this point, energy dissipation will continue as long as the primary rotates asynchronously. However, there will not be enough energy to allow for a complex spin state in the secondary. As dissipation continues within the primary, the separation distance $r$ will increase while the primary's spin rate $\omega_A$ will decrease.

The primary methods of energy dissipation considered are tidal torques. Starting from the classical tidal torque equation in \cite{murray2000solar}, \cite{meyer2023energy} generalized this to three dimensions as:
\begin{equation}
    \vec{\Gamma}_B = \frac{3GM_A^2M_B^2}{2r^6R_B}\left(\frac{3}{4\pi\rho}\right)^2\frac{k_B}{Q_B}\left(\frac{-(\dot{\vec{\phi}}-(\dot{\vec{\phi}}\cdot\hat{r})\hat{r})}{|\dot{\vec{\phi}}-(\dot{\vec{\phi}}\cdot\hat{r})\hat{r}|}\right)
\end{equation}
where
\begin{equation}
    \dot{\vec{\phi}}=\vec{\omega}_B-\vec{\omega}_{orbit}.
\end{equation}
The same equation can be applied to the primary, and the torque on the orbit is equal and opposite to the sum of the torques on the primary and secondary. As derived in \cite{meyer2023energy}, this adds an acceleration to the orbit:
\begin{equation}
    \ddot{\vec{r}}=\frac{\vec{\Gamma}_{orbit}\times\vec{r}}{Mr^2}.
\end{equation}
where the torque on the orbit is simply $\vec{\Gamma}_{orbit}=-(\vec{\Gamma}_{A}+\vec{\Gamma}_{B})$ to conserve angular momentum in the system.

For the ratio of tidal quality factor to Love number, we initially use the radius-dependency derived by \cite{jacobson2011long}. This was shown to be a good approximation for rubble-pile asteroids \citep{nimmo2019tidal,pou2024tidal}, and is given as
 \begin{equation}
     \frac{Q_i}{k_i}\approx300R_i
 \end{equation}
for body $i$, where $R_i$ is in meters. However, there is significant uncertainty around these values, and in reality could be larger or smaller than this value. Nominally, this gives us $Q_A/k_A = 180,000$ for the primary and $Q_B/k_B = 67500$ for the secondary.

\begin{figure}[ht!]
   \centering
   \includegraphics[width = 3in]{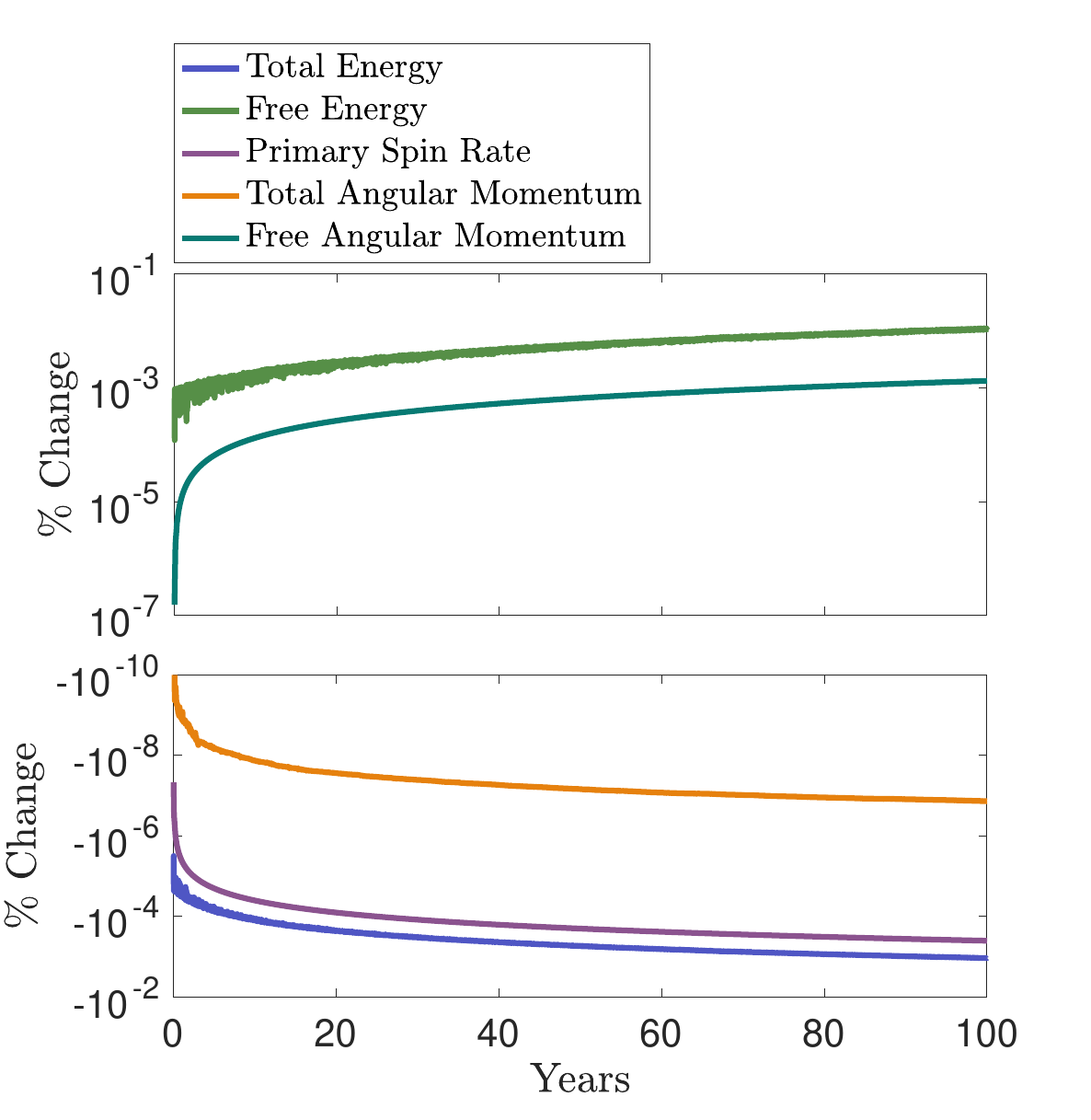} 
   \caption{The percent change in total energy, free energy, primary spin rate, total angular momentum, and free angular momentum. We split between increasing (to plot) and decreasing (bottom plot) quantities on a logarithmic scale.}
   \label{fig:dissipation}
\end{figure}

Using the results from Section \ref{sec:example}, we integrate the equations of motion forward for another 100 years using the sphere-ellipsoid equations of motion. Because of the asynchronous rotation of the secondary, we do not need to include the BYORP effect in these simulations. The percent change in total energy, free energy, and primary spin rate in these units are plotted in Fig. \ref{fig:dissipation}, demonstrating the energy dissipation. Here we are defining free energy as the total energy minus the contribution from the primary.

In Fig. \ref{fig:dissipation}, we also plot the total and free angular momentum. This ensures our system is conserving total angular momentum accurate to within $10^{-6}$ percent. However, due to the secular evolution, the free angular momentum (the contribution of the secondary and the orbit) is increasing.

\begin{figure}[ht!]
   \centering
   \includegraphics[width = 3in]{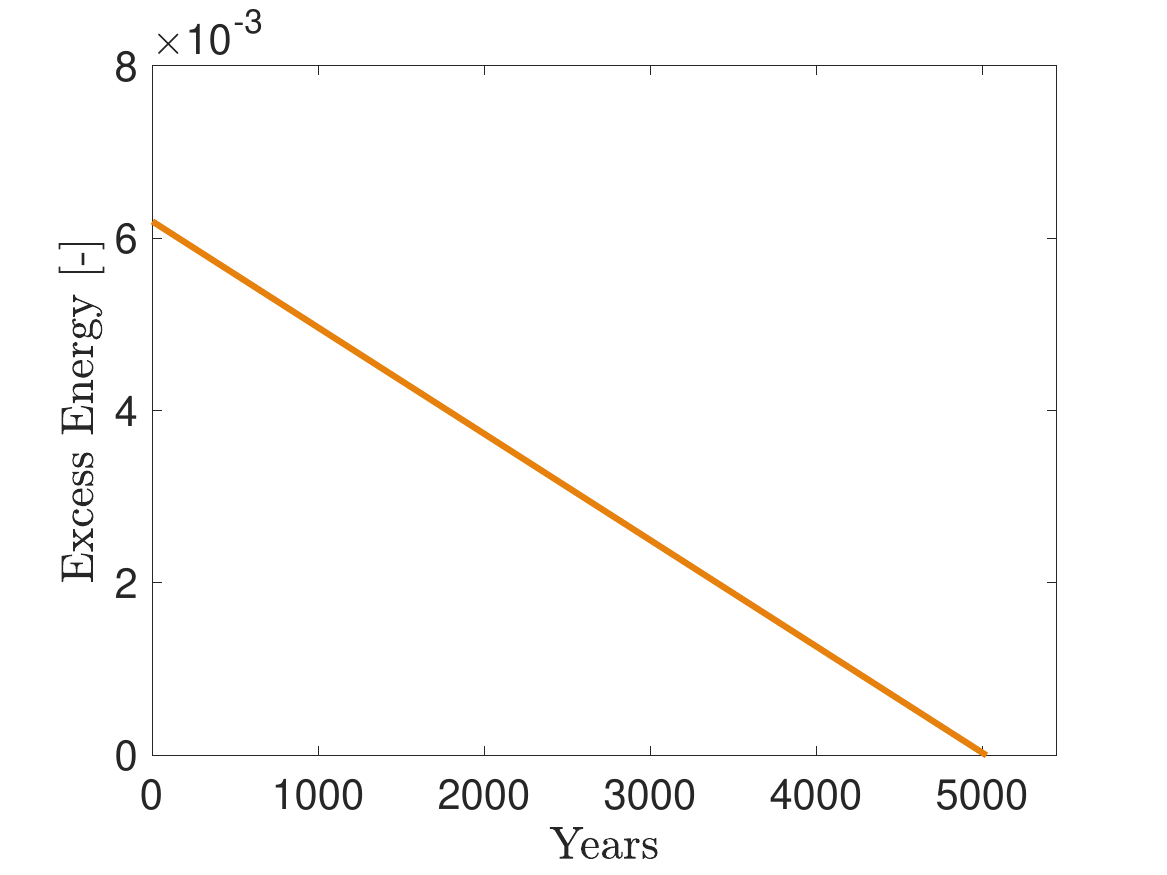} 
   \caption{The excess energy of the system at this level of angular momentum. When the curve reaches zero, there is no longer any excess energy and the system must be in the singly-synchronous configuration.}
   \label{fig:excess_energy}
\end{figure}

To predict the system's future behavior, we perform a linear fit to both the total energy and the primary spin rate. The linear fit is a conservative estimate, as dissipation will in reality slow as the separation distance increases. Using this fit, we can extrapolate the total energy and the primary's spin rate at a future time. Using the primary's spin rate, numerically solving Eq. \ref{eq:H} provides the separation distance required to conserve angular momentum, then Eq. \ref{eq:E} at these values gives the energy for the singly-synchronous configuration at this angular momentum level, what we call the minimum energy. The difference between the total energy and this minimum energy is the excess energy, which is plotted in Fig. \ref{fig:excess_energy}. 

As we see, the excess energy is greater than 0 for around $5,000$ years in this case, suggesting NPA rotation would in fact be relatively short lived for these tidal parameters. Indeed, a close Earth encounter for 1991VH did not occur within the past $12,000$ years \citep{fuentes2022semi}.

\begin{figure}[ht!]
   \centering
   \includegraphics[width = 3in]{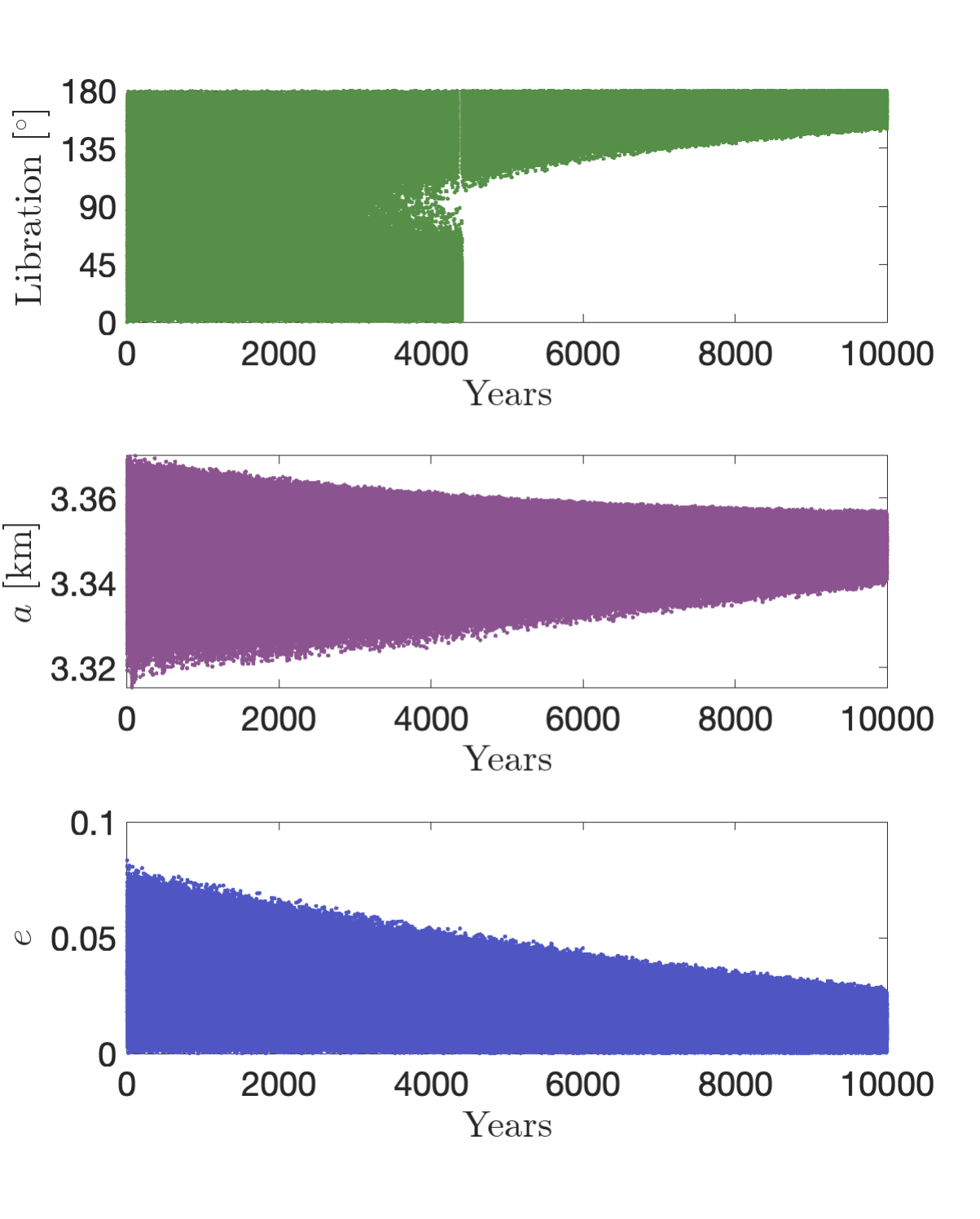} 
   \caption{The libration angle (top), semimajor axis (middle), and eccentricity (bottom) of our example system undergoing tidal dissipation.}
   \label{fig:dissipation_orbit}
\end{figure}

To validate this predicted time for NPA rotation, we integrate the system for $10,000$ years. The results of this integration are shown in Fig. \ref{fig:dissipation_orbit}, plotting the libration angle, semimajor axis, and eccentricity. In this plot, the libration angle is the angle between the secondary's body-fixed $x$-axis and the position vector of the secondary relative to the primary.

From Fig. \ref{fig:dissipation_orbit}, we see very good agreement with the predicted duration of NPA rotation and these fully numerical results. The libration angle stops circulation a little before $5,000$ years, very similar to the prediction by our linear extrapolation. In this example, the secondary settles into the anti-synchronous case where it has flipped $180^\circ$ from its initial orientation. This plot also shows the reduction of eccentricity, which happens much faster than predicted in analytical models \citep{taylor2011binary,meyer2023energy}. During this initial energy dissipation, the eccentricity is dissipating quicker than the semimajor axis is expanding, although we do see a secular increasing trend in the semimajor axis envelope toward the end of the simulation.

With the linear dissipation model validated by this detailed numerical integration, we now extend the analysis to different values of secondary tidal parameters. The secondary's libration dissipation is driven primarily by the parameters of the secondary \citep{murray2000solar,meyer2023energy}, so we can limit our analysis to only these parameters while keeping the primary's parameters constant.

\begin{figure}[ht!]
   \centering
   \includegraphics[width = 3in]{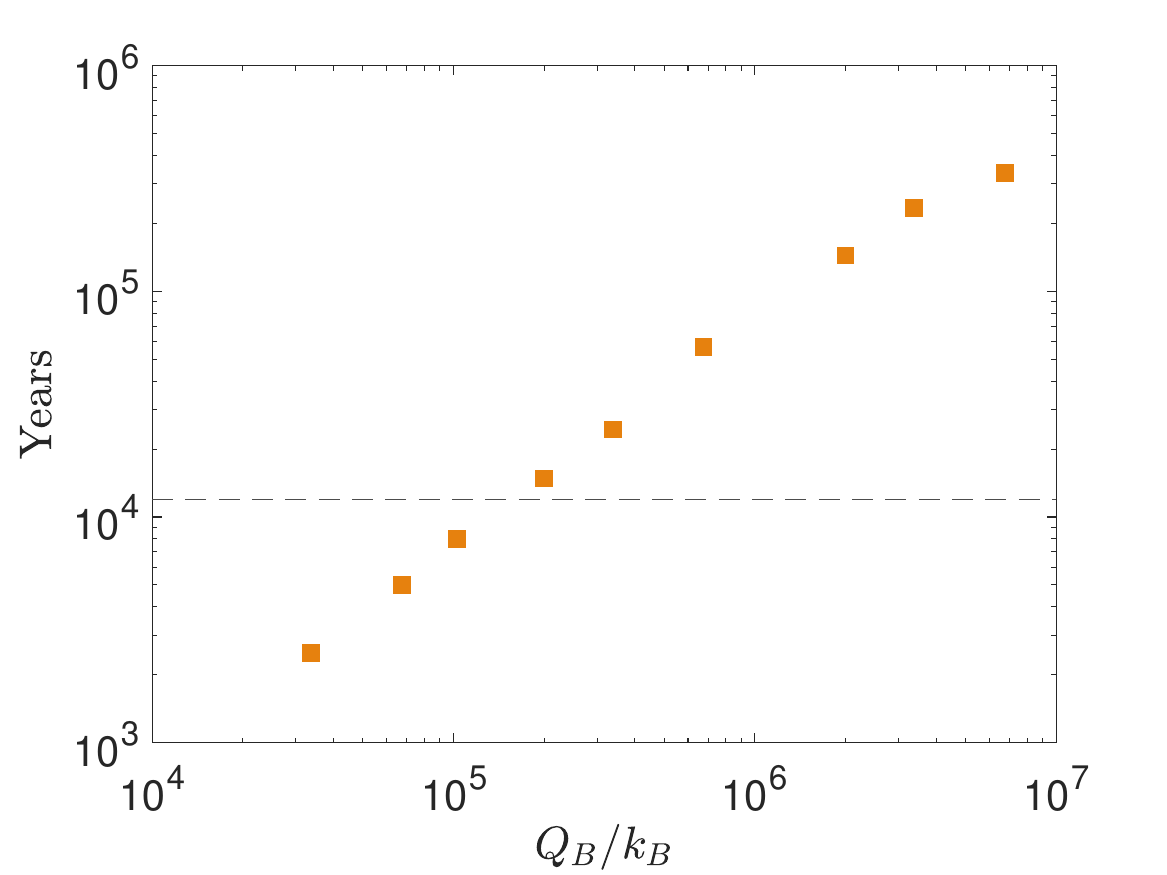} 
   \caption{The permissible time of NPA rotation as a function of secondary tidal parameters. Close Earth encounters were possible starting around 12,000 years ago, which is marked as a dashed black line. This places a minimum boundary on secondary tidal parameters assuming 1991VH was excited by a past Earth encounter.}
   \label{fig:secondary_parameters}
\end{figure}

Fig. \ref{fig:secondary_parameters} plots the total time for which NPA rotation is permissible as a function of secondary tidal parameters, keeping the tidal parameters of the primary constant. The latest a flyby could have occurred was around $12,000$ years ago, which is plotted as a dashed black line. Assuming it was an Earth encounter that provided the excitation to place 1991VH in an excited dynamical state, this places a lower bound on $Q_B/k_B$ of around $2\times10^5$, although in reality it would likely be higher. This is relatively large for a rubble pile secondary \citep{pou2024tidal}, and using the model of \cite{nimmo2019tidal} would require a 15 m or smaller dissipative regolith layer or less around the secondary. However, these values are still within the realm of possibility.

\section{Discussion} \label{sec:discussion}
In this work we have investigated the feasibility that the current excited and chaotic dynamical state of 1991VH is the result of a previous close encounter with Earth. We find that it is relatively easy to excite NPA rotation of the secondary of 1991VH, and this NPA rotation is generally chaotic and can cross between planar spin-orbit resonances thanks to the addition of two out-of-plane rotational degrees of freedom. We find it is most likely that 1991VH was in a singly-synchronous configuration with an orbit period between 28 and 35 hours, although probably not within the 30-33 hour range. A single close Earth encounter around between $50,000$ km and $80,000$ km could have then placed the system into its current dynamical state, slightly closer but still in line with our initial analytic prediction. Because the current eccentricity of 1991VH is relatively small, this indicates the semimajor axis and orbit period of the system did not significantly change as a result of the flyby. However, the spin-orbit coupled dynamics are sufficient to place the secondary into its currently excited rotation state as a result of this flyby. A flyby closer than $50,000$ km is possible if secular dissipation has reduced the eccentricity into the range observed today.

Using only the internal spin-orbit coupled dynamics characteristic of binary asteroids, we find a possible range of close encounter geometries can excite the system to the level observed today. We emphasize this requires only a single external perturbation, and the internal system dynamics naturally evolve the system into this state.  However, there is a large range of possible post-encounter dynamical states, and the number of systems that reproduce the observed secondary rotation period is the minority. One point of consideration is that 1991VH itself is a minority among the binary asteroid population, in that it exists in this excited, asynchronous state \citep{pravec2016binary}. In our simulations, we see most flybys result in a system remaining within the synchronous configuration, consistent with the observed binary asteroid population. The second point of consideration is that our simulations are not calculating the probability that 1991VH experienced this Earth encounter. This probability was calculated in \cite{fuentes2022semi}, who find a $50\%$ chance 1991VH was significantly perturbed by a close encounter within the past $250,000$ years. Instead, our simulations simply show that if 1991VH experienced such an Earth encounter, this could explain its current dynamical state.

Finally, in reality, a wider range of flyby geometries is possible, as we have considered only a single flyby whereas in reality there the system would experience successive close approaches over time. These successive close encounters, along with energy dissipation, would further evolve the system and could be responsible for the observed state of the system today.

One interesting point of discussion is how long-lived we would expect this state of NPA rotation to last. In our analysis, we only include dissipation caused by tidal torques, while in reality there are many mechanisms of energy dissipation, including NPA rotation \citep{vokrouhlicky2007generalized}, tidal saltation \citep{harris2009shapes}, YORP \citep{rubincam2000radiative}, and surface motion \citep{agrusa2022rotation}. While these could increase the rate of tidal dissipation, tidal torque is the strongest among these \citep{meyer2023energy}, and these other dissipation mechanisms could in effect be wrapped within the uncertainty on the tidal $Q/k$ parameters. These additional dissipation mechanisms could be thought of as simply increasing the rate of dissipation used in this work. This simplification is adopted to increase numerical stability for our long-term simulations. Nominally, the proximity and rubble-pile nature of binary asteroids suggests these systems can re-synchronize quite quickly \citep{meyer2022modeling}. This is contrary to the observed state of 1991VH, as the system has not undergone a close planetary encounter within the past $12,000$ years \citep{fuentes2022semi}. This constraint places a lower bound on the secondary's $Q_B/k_B$ tidal parameter: $Q_B/k_B\gtrsim2\times10^5$. This is consistent with recent results from \cite{pou2024tidal}, who place a general upper bound for $Q_B/k_B$ roughly around $\lesssim1\times10^6$, moderately higher than our minimum constraint. However, their work also uses the analytical approach to eccentricity damping \citep{taylor2011binary,boue2019tidal,murray2000solar}, which was shown in \cite{meyer2023energy} to generally over-estimate the eccentricity damping time in binary asteroids. Thus, while our resulting $Q_B/k_B$ is slightly larger than typically found, the required values of $Q_B/k_B$ in 1991VH are in general consistent with the current literature. Another consideration is successive close encounters would counteract energy dissipation by providing additional perturbations over time. This means our calculated lower bound of $Q_B/k_B$ is a conservative estimate. 

When considering only binary asteroids that have estimates for both the orbit period and secondary rotation period, we find around $15\%$ of these systems have asynchronous secondaries. If we assume these systems became asynchronous as a result of close planetary encounters, this potentially provides information about energy dissipation rates within these systems. In their work, \cite{fuentes2022semi} calculate the probability of a close planetary encounter for two binary asteroids: (35107) 1991VH and (175706) 1996FG3. For 1991VH, they found a likelihood of $15\%$ that this system experienced a close encounter within the past $70,000$ years. For 1996FG3, they found the same likelihood occurs around $30,000$ years ago. These are of the same order of magnitude, so if we assume these numbers can be generalized to the population of NEA binary asteroids at large, then $15\%$ of NEA binary asteroids would be perturbed by these close flybys on the order of $10,000$ to $100,000$ years ago. If we take 1991VH to be a characteristic binary binary asteroid, this would place the tidal parameters of secondaries $Q_B/k_B$ roughly around $1\times10^5$ to $1\times10^6$ in order to maintain around $15\%$ of binary asteroids to be asynchronous. However, far more data and analysis is needed to substantiate this argument.

Due to the orbital evolution of NEAs, we argue a planetary encounter is the most likely source for the excited dynamics in 1991VH. However, many other possible explanations exist. For example, one or several impacts onto the secondary could induce these unstable dynamics. This was demonstrated by the DART impact, which induced NPA rotation in the secondary of (65803) Didymos \citep{pravec2024rotational}. However, an NEA the size of the secondary of 1991VH has an extremely small probability of experiencing a significant collision in the past million years (on the order of $10^{-13}$) \citep{bottke1994collisional}. As discussed, it is unlikely that NPA rotation persists longer than this time for 1991VH. Another possible explanation is if 1991VH was previously in a BYORP-tide equilibrium, and a resonance between the orbit and primary spin destabilized the rotation of the secondary \citep{wang2021break}. If the secondary shape is near a resonance between the natural frequencies within the system, this destabilization would result in NPA rotation. However, the rapid rotation of the primary in 1991VH means the system is far from any low-order resonance between its spin and the mean motion. Alternatively, BYORP and tidal evolution could have migrated the system into a wider configuration where the secondary's rotation was no longer synchronized by the primary \citep{jacobson2013formation}. However, this scenario would then require a mechanism to reduce the semimajor axis. Another cause of the excited dynamics could be if 1991VH was previously a triple system whose secondaries merged, resulting in an unstable secondary rotational state. The very elongated shape of the secondary in 1991VH predicted by radar data could be explained by such a contact binary \citep{naidu2015radar}, although this evidence is circumstantial at best, and there is no concrete evidence of a contact binary satellite. The contact binary secondary of (152830) Dinkinesh could be an example of such a merger, although its satellite is much more elongated than even the upper estimates for 1991VH \citep{levison2024contact}. Furthermore, the merger would have had to occur relatively recently for the excited rotational state to persist to today, and the probability of such an event is unknown.  All of these scenarios could be examined in further detail. However, the high probability of recent and close encounters with the terrestrial planets suggests this is the most likely cause for the unstable dynamics \citep{fuentes2022semi}. While we have exclusively focused on Earth encounters in this work, encounters with Mars are also possible.

In this work we have demonstrated one possible cause of the excited dynamical state of 1991VH. In this approach, a single Earth flyby within the past $\sim100,000$ years is capable of producing the currently observed state in the system. This state could then persist to the current day without energy dissipation returning it to the archetypal singly-synchronous state, provided the secondary is not very efficient at energy dissipation or repeated close encounters counteract energy dissipation.

\section*{Acknowledgements}
A.J.M. acknowledges support from the Planetary Defense Conference Student Grant. A portion of this work was conducted at the Jet Propulsion Laboratory, California Institute of Technology, under a contract with the National Aeronautics and Space Administration (80NM0018D0004). The work by P.P. was supported by the Grant Agency of the Czech Republic, Grant 23-04946S.

\bibliography{bib}{}

\begin{thebibliography}{}
\expandafter\ifx\csname natexlab\endcsname\relax\def\natexlab#1{#1}\fi
\providecommand{\url}[1]{\href{#1}{#1}}
\providecommand{\dodoi}[1]{doi:~\href{http://doi.org/#1}{\nolinkurl{#1}}}
\providecommand{\doeprint}[1]{\href{http://ascl.net/#1}{\nolinkurl{http://ascl.net/#1}}}
\providecommand{\doarXiv}[1]{\href{https://arxiv.org/abs/#1}{\nolinkurl{https://arxiv.org/abs/#1}}}

\bibitem[{Agrusa {et~al.}(2022)Agrusa, Ballouz, Meyer, Tasev, Noiset, Karatekin, Michel, Richardson, \& Hirabayashi}]{agrusa2022rotation}
Agrusa, H., Ballouz, R., Meyer, A.~J., {et~al.} 2022, Astronomy \& Astrophysics, 664, L3

\bibitem[{Agrusa {et~al.}(2021)Agrusa, Gkolias, Tsiganis, Richardson, Meyer, Scheeres, {\'C}uk, Jacobson, Michel, Karatekin, {et~al.}}]{agrusa2021excited}
Agrusa, H.~F., Gkolias, I., Tsiganis, K., {et~al.} 2021, Icarus, 370, 114624

\bibitem[{Bottke {et~al.}(1994)Bottke, Nolan, Greenberg, \& Kolvoord}]{bottke1994collisional}
Bottke, W.~F., Nolan, M.~C., Greenberg, R., \& Kolvoord, R.~A. 1994, Hazards due to comets and asteroids, 337

\bibitem[{Bou{\'e} \& Efroimsky(2019)}]{boue2019tidal}
Bou{\'e}, G., \& Efroimsky, M. 2019, Celestial Mechanics and Dynamical Astronomy, 131, 1

\bibitem[{Cueva {et~al.}(2024)Cueva, McMahon, Meyer, Scheeres, Hirabayashi, Raducan, Jacobson, \& Merrill}]{cueva2024secular}
Cueva, R.~H., McMahon, J.~W., Meyer, A.~J., {et~al.} 2024, The Planetary Science Journal, 5, 48

\bibitem[{{\'C}uk \& Burns(2005)}]{cuk2005effects}
{\'C}uk, M., \& Burns, J.~A. 2005, Icarus, 176, 418

\bibitem[{{\'C}uk {et~al.}(2021){\'C}uk, Jacobson, \& Walsh}]{cuk2021barrel}
{\'C}uk, M., Jacobson, S.~A., \& Walsh, K.~J. 2021, The Planetary Science Journal, 2, 231

\bibitem[{Davis \& Scheeres(2020)}]{davis2020doubly}
Davis, A.~B., \& Scheeres, D.~J. 2020, Icarus, 341, 113439

\bibitem[{Davis \& Scheeres(2021)}]{davis2021gubas}
---. 2021, Astrophysics Source Code Library, ascl

\bibitem[{Fang \& Margot(2011)}]{fang2011binary}
Fang, J., \& Margot, J.-L. 2011, The Astronomical Journal, 143, 25

\bibitem[{Fuentes-Mu{\~n}oz {et~al.}(2022)Fuentes-Mu{\~n}oz, Meyer, \& Scheeres}]{fuentes2022semi}
Fuentes-Mu{\~n}oz, O., Meyer, A.~J., \& Scheeres, D.~J. 2022, The Planetary Science Journal, 3, 257

\bibitem[{Fuentes-Muñoz {et~al.}(2024)Fuentes-Muñoz, Farnocchia, Naidu, \& Park}]{fuentes2024fpr}
Fuentes-Muñoz, O., Farnocchia, D., Naidu, S.~P., \& Park, R.~S. 2024, The Astronomical Journal, 167, 290, \dodoi{10.3847/1538-3881/ad4291}

\bibitem[{{Gaia Collaboration} {et~al.}(2023){Gaia Collaboration}, {David, P.}, {Mignard, F.}, {Hestroffer, D.}, {Tanga, P.}, {Spoto, F.}, {Berthier, J.}, {Pauwels, T.}, {Roux, W.}, {Barbier, A.}, {Cellino, A.}, {Carry, B.}, {Delbo, M.}, {Dell’Oro, A.}, {Fouron, C.}, {Galluccio, L.}, {Klioner, S. A.}, {Mary, N.}, {Muinonen, K.}, {Ordenovic, C.}, {Oreshina-Slezak, I.}, {Panem, C.}, {Petit, J.-M.}, {Portell, J.}, {Brown, A. G. A.}, {Thuillot, W.}, {Vallenari, A.}, {Prusti, T.}, {de Bruijne, J. H. J.}, {Arenou, F.}, {Babusiaux, C.}, {Biermann, M.}, {Creevey, O. L.}, {Ducourant, C.}, {Evans, D. W.}, {Eyer, L.}, {Guerra, R.}, {Hutton, A.}, {Jordi, C.}, {Lammers, U.}, {Lindegren, L.}, {Luri, X.}, {Randich, S.}, {Sartoretti, P.}, {Smiljanic, R.}, {Walton, N. A.}, {Bailer-Jones, C. A. L.}, {Bastian, U.}, {Cropper, M.}, {Drimmel, R.}, {Katz, D.}, {Soubiran, C.}, {van Leeuwen, F.}, {Audard, M.}, {Bakker, J.}, {Blomme, R.}, {Castañeda, J.}, {De Angeli, F.}, {Fabricius, C.}, {Fouesneau, M.}, {Frémat, Y.}, {Guerrier,
  A.}, {Masana, E.}, {Messineo, R.}, {Nicolas, C.}, {Nienartowicz, K.}, {Pailler, F.}, {Panuzzo, P.}, {Riclet, F.}, {Seabroke, G. M.}, {Sordo, R.}, {Thévenin, F.}, {Gracia-Abril, G.}, {Teyssier, D.}, {Altmann, M.}, {Benson, K.}, {Burgess, P. W.}, {Busonero, D.}, {Busso, G.}, {Cánovas, H.}, {Cheek, N.}, {Clementini, G.}, {Damerdji, Y.}, {Davidson, M.}, {de Teodoro, P.}, {Delchambre, L.}, {Fraile Garcia, E.}, {Garabato, D.}, {García-Lario, P.}, {Garralda Torres, N.}, {Gavras, P.}, {Haigron, R.}, {Hambly, N. C.}, {Harrison, D. L.}, {Hatzidimitriou, D.}, {Hernández, J.}, {Hodgkin, S. T.}, {Holl, B.}, {Jamal, S.}, {Jordan, S.}, {Krone-Martins, A.}, {Lanzafame, A. C.}, {Löffler, W.}, {Lorca, A.}, {Marchal, O.}, {Marrese, P. M.}, {Moitinho, A.}, {Nuñez Campos, M.}, {Osborne, P.}, {Pancino, E.}, {Recio-Blanco, A.}, {Riello, M.}, {Rimoldini, L.}, {Robin, A. C.}, {Roegiers, T.}, {Sarro, L. M.}, {Schultheis, M.}, {Siopis, C.}, {Smith, M.}, {Sozzetti, A.}, {Utrilla, E.}, {van Leeuwen, M.}, {Weingrill, K.}, {Abbas,
  U.}, {Ábrahám, P.}, {Abreu Aramburu, A.}, {Aerts, C.}, {Altavilla, G.}, {Álvarez, M. A.}, {Alves, J.}, {Anderson, R. I.}, {Antoja, T.}, {Baines, D.}, {Baker, S. G.}, {Balog, Z.}, {Barache, C.}, {Barbato, D.}, {Barros, M.}, {Barstow, M. A.}, {Bartolomé, S.}, {Bashi, D.}, {Bauchet, N.}, {Baudeau, N.}, {Becciani, U.}, {Bedin, L. R.}, {Bellas-Velidis, I.}, {Bellazzini, M.}, {Beordo, W.}, {Berihuete, A.}, {Bernet, M.}, {Bertolotto, C.}, {Bertone, S.}, {Bianchi, L.}, {Binnenfeld, A.}, {Blazere, A.}, {Boch, T.}, {Bombrun, A.}, {Bouquillon, S.}, {Bragaglia, A.}, {Braine, J.}, {Bramante, L.}, {Breedt, E.}, {Bressan, A.}, {Brouillet, N.}, {Brugaletta, E.}, {Bucciarelli, B.}, {Butkevich, A. G.}, {Buzzi, R.}, {Caffau, E.}, {Cancelliere, R.}, {Cannizzo, S.}, {Carballo, R.}, {Carlucci, T.}, {Carnerero, M. I.}, {Carrasco, J. M.}, {Carretero, J.}, {Carton, S.}, {Casamiquela, L.}, {Castellani, M.}, {Castro-Ginard, A.}, {Cesare, V.}, {Charlot, P.}, {Chemin, L.}, {Chiaramida, V.}, {Chiavassa, A.}, {Chornay, N.}, {Collins,
  R.}, {Contursi, G.}, {Cooper, W. J.}, {Cornez, T.}, {Crosta, M.}, {Crowley, C.}, {Dafonte, C.}, {de Laverny, P.}, {De Luise, F.}, {De March, R.}, {de Souza, R.}, {de Torres, A.}, {del Peloso, E. F.}, {Delgado, A.}, {Dharmawardena, T. E.}, {Diakite, S.}, {Diener, C.}, {Distefano, E.}, {Dolding, C.}, {Dsilva, K.}, {Durán, J.}, {Enke, H.}, {Esquej, P.}, {Fabre, C.}, {Fabrizio, M.}, {Faigler, S.}, {Fatović, M.}, {Fedorets, G.}, {Fernández-Hernández, J.}, {Fernique, P.}, {Figueras, F.}, {Fournier, Y.}, {Gai, M.}, {Galinier, M.}, {Garcia-Gutierrez, A.}, {García-Torres, M.}, {Garofalo, A.}, {Gerlach, E.}, {Geyer, R.}, {Giacobbe, P.}, {Gilmore, G.}, {Girona, S.}, {Giuffrida, G.}, {Gomel, R.}, {Gomez, A.}, {González-Núñez, J.}, {González-Santamaría, I.}, {Gosset, E.}, {Granvik, M.}, {Gregori Barrera, V.}, {Gutiérrez-Sánchez, R.}, {Haywood, M.}, {Helmer, A.}, {Helmi, A.}, {Henares, K.}, {Hidalgo, S. L.}, {Hilger, T.}, {Hobbs, D.}, {Hottier, C.}, {Huckle, H. E.}, {Jabłońska, M.}, {Jansen, F.},
  {Jiménez-Arranz, Ó.}, {Juaristi Campillo, J.}, {Khanna, S.}, {Kordopatis, G.}, {Kóspál, Á.}, {Kostrzewa-Rutkowska, Z.}, {Kun, M.}, {Lambert, S.}, {Lanza, A. F.}, {Le Campion, J.-F.}, {Lebreton, Y.}, {Lebzelter, T.}, {Leccia, S.}, {Lecoeur-Taibi, I.}, {Lecoutre, G.}, {Liao, S.}, {Liberato, L.}, {Licata, E.}, {Lindstrøm, H. E. P.}, {Lister, T. A.}, {Livanou, E.}, {Lobel, A.}, {Loup, C.}, {Mahy, L.}, {Mann, R. G.}, {Manteiga, M.}, {Marchant, J. M.}, {Marconi, M.}, {Marín Pina, D.}, {Marinoni, S.}, {Marshall, D. J.}, {Martín Lozano, J.}, {Martín-Fleitas, J. M.}, {Marton, G.}, {Masip, A.}, {Massari, D.}, {Mastrobuono-Battisti, A.}, {Mazeh, T.}, {McMillan, P. J.}, {Meichsner, J.}, {Messina, S.}, {Michalik, D.}, {Millar, N. R.}, {Mints, A.}, {Molina, D.}, {Molinaro, R.}, {Molnár, L.}, {Monari, G.}, {Monguió, M.}, {Montegriffo, P.}, {Montero, A.}, {Mor, R.}, {Mora, A.}, {Morbidelli, R.}, {Morel, T.}, {Morris, D.}, {Mowlavi, N.}, {Munoz, D.}, {Muraveva, T.}, {Murphy, C. P.}, {Musella, I.}, {Nagy, Z.},
  {Nieto, S.}, {Noval, L.}, {Ogden, A.}, {Pagani, C.}, {Pagano, I.}, {Palaversa, L.}, {Palicio, P. A.}, {Pallas-Quintela, L.}, {Panahi, A.}, {Payne-Wardenaar, S.}, {Pegoraro, L.}, {Penttilä, A.}, {Pesciullesi, P.}, {Piersimoni, A. M.}, {Pinamonti, M.}, {Pineau, F.-X.}, {Plachy, E.}, {Plum, G.}, {Poggio, E.}, {Pourbaix, D.}, {Prša, A.}, {Pulone, L.}, {Racero, E.}, {Rainer, M.}, {Raiteri, C. M.}, {Ramos, P.}, {Ramos-Lerate, M.}, {Ratajczak, M.}, {Re Fiorentin, P.}, {Regibo, S.}, {Reylé, C.}, {Ripepi, V.}, {Riva, A.}, {Rix, H.-W.}, {Rixon, G.}, {Robichon, N.}, {Robin, C.}, {Romero-Gómez, M.}, {Rowell, N.}, {Royer, F.}, {Ruz Mieres, D.}, {Rybicki, K. A.}, {Sadowski, G.}, {Sáez Núñez, A.}, {Sagristà Sellés, A.}, {Sahlmann, J.}, {Sanchez Gimenez, V.}, {Sanna, N.}, {Santoveña, R.}, {Sarasso, M.}, {Sarrate Riera, C.}, {Sciacca, E.}, {Segovia, J. C.}, {Ségransan, D.}, {Shahaf, S.}, {Siebert, A.}, {Siltala, L.}, {Slezak, E.}, {Smart, R. L.}, {Snaith, O. N.}, {Solano, E.}, {Solitro, F.}, {Souami, D.},
  {Souchay, J.}, {Spina, L.}, {Spitoni, E.}, {Squillante, L. A.}, {Steele, I. A.}, {Steidelmüller, H.}, {Surdej, J.}, {Szabados, L.}, {Taris, F.}, {Taylor, M. B.}, {Teixeira, R.}, {Tisanić, K.}, {Tolomei, L.}, {Torra, F.}, {Torralba Elipe, G.}, {Trabucchi, M.}, {Tsantaki, M.}, {Ulla, A.}, {Unger, N.}, {Vanel, O.}, {Vecchiato, A.}, {Vicente, D.}, {Voutsinas, S.}, {Weiler, M.}, {Wyrzykowski, Ł.}, {Zhao, H.}, {Zorec, J.}, {Zwitter, T.}, {Balaguer-Núñez, L.}, {Leclerc, N.}, {Morgenthaler, S.}, {Robert, G.}, \& {Zucker, S.}}]{david2023gaia}
{Gaia Collaboration}, {David, P.}, {Mignard, F.}, {et~al.} 2023, \aap, 680, A37, \dodoi{10.1051/0004-6361/202347270}

\bibitem[{Gottlieb(1993)}]{gottlieb1993fast}
Gottlieb, R.~G. 1993, Fast gravity, gravity partials, normalized gravity, gravity gradient torque and magnetic field: derivation, code and data, Tech. rep.

\bibitem[{Harris {et~al.}(2009)Harris, Fahnestock, \& Pravec}]{harris2009shapes}
Harris, A.~W., Fahnestock, E.~G., \& Pravec, P. 2009, Icarus, 199, 310

\bibitem[{Jacobson \& Scheeres(2011)}]{jacobson2011long}
Jacobson, S.~A., \& Scheeres, D.~J. 2011, The Astrophysical Journal Letters, 736, L19

\bibitem[{Jacobson {et~al.}(2013)Jacobson, Scheeres, \& McMahon}]{jacobson2013formation}
Jacobson, S.~A., Scheeres, D.~J., \& McMahon, J. 2013, The Astrophysical Journal, 780, 60

\bibitem[{Levison {et~al.}(2024)Levison, Marchi, Noll, Spencer, Statler, Bell~III, Bierhaus, Binzel, Bottke, Britt, {et~al.}}]{levison2024contact}
Levison, H.~F., Marchi, S., Noll, K.~S., {et~al.} 2024, Nature, 629, 1015

\bibitem[{Liberato {et~al.}(2023)Liberato, Ara{\'u}jo, \& Winter}]{liberato2023known}
Liberato, L., Ara{\'u}jo, R., \& Winter, O. 2023, The European Physical Journal Special Topics, 232, 3007

\bibitem[{Maciejewski(1995)}]{maciejewski1995reduction}
Maciejewski, A.~J. 1995, Celestial Mechanics and Dynamical Astronomy, 63, 1

\bibitem[{Merrill {et~al.}(2024)Merrill, Kubas, Meyer, \& Raducan}]{merrill2024age}
Merrill, C., Kubas, A., Meyer, A., \& Raducan, S. 2024, Astronomy \& Astrophysics, 684, L20

\bibitem[{Meyer {et~al.}(2022{\natexlab{a}})Meyer, Gkolias, Tsiganis, Scheeres, Naidu, Pravec, \& Benson}]{meyer2022chaotic}
Meyer, A., Gkolias, I., Tsiganis, K., {et~al.} 2022{\natexlab{a}}, in AAS/Division for Planetary Sciences Meeting Abstracts, Vol.~54, 202--05

\bibitem[{Meyer {et~al.}(2022{\natexlab{b}})Meyer, Scheeres, \& Benson}]{meyer2022modeling}
Meyer, A., Scheeres, D., \& Benson, C. 2022{\natexlab{b}}, in AAS/Division of Dynamical Astronomy Meeting, Vol.~54, 200--03

\bibitem[{Meyer \& Scheeres(2021)}]{meyer2021effect}
Meyer, A.~J., \& Scheeres, D.~J. 2021, Icarus, 367, 114554

\bibitem[{Meyer {et~al.}(2023{\natexlab{a}})Meyer, Scheeres, Agrusa, Noiset, McMahon, Karatekin, Hirabayashi, \& Nakano}]{meyer2023energy}
Meyer, A.~J., Scheeres, D.~J., Agrusa, H.~F., {et~al.} 2023{\natexlab{a}}, Icarus, 391, 115323

\bibitem[{Meyer {et~al.}(2021)Meyer, Gkolias, Gaitanas, Agrusa, Scheeres, Tsiganis, Pravec, Benner, Ferrari, \& Michel}]{meyer2021libration}
Meyer, A.~J., Gkolias, I., Gaitanas, M., {et~al.} 2021, The planetary science journal, 2, 242

\bibitem[{Meyer {et~al.}(2023{\natexlab{b}})Meyer, Agrusa, Richardson, Daly, Fuentes-Mu{\~n}oz, Hirabayashi, Michel, Merrill, Nakano, Cheng, {et~al.}}]{meyer2023perturbed}
Meyer, A.~J., Agrusa, H.~F., Richardson, D.~C., {et~al.} 2023{\natexlab{b}}, The Planetary Science Journal, 4, 141

\bibitem[{Murray \& Dermott(2000)}]{murray2000solar}
Murray, C.~D., \& Dermott, S.~F. 2000, Solar system dynamics (Cambridge university press)

\bibitem[{Naidu {et~al.}(2018)Naidu, Margot, Benner, Taylor, Nolan, Magri, Brozovic, Busch, \& Giorgini}]{naidu2018radar}
Naidu, S., Margot, J.-L., Benner, L., {et~al.} 2018, in AAS/Division for Planetary Sciences Meeting Abstracts\# 50, Vol.~50, 312--09

\bibitem[{Naidu \& Margot(2015)}]{naidu2015near}
Naidu, S.~P., \& Margot, J.-L. 2015, The Astronomical Journal, 149, 80

\bibitem[{Naidu {et~al.}(2012)Naidu, Margot, Busch, Taylor, Nolan, Howell, Giorgini, Benner, Brozovic, \& Magri}]{naidu2012dynamics}
Naidu, S.~P., Margot, J., Busch, M., {et~al.} 2012, in AAS/Division of Dynamical Astronomy Meeting\# 43, Vol.~43, 7--07

\bibitem[{Naidu {et~al.}(2015)Naidu, Margot, Taylor, Nolan, Busch, Benner, Brozovic, Giorgini, Jao, \& Magri}]{naidu2015radar}
Naidu, S.~P., Margot, J.-L., Taylor, P.~A., {et~al.} 2015, The Astronomical Journal, 150, 54

\bibitem[{Nimmo \& Matsuyama(2019)}]{nimmo2019tidal}
Nimmo, F., \& Matsuyama, I. 2019, Icarus, 321, 715

\bibitem[{Nugent {et~al.}(2016)Nugent, Mainzer, Bauer, Cutri, Kramer, Grav, Masiero, Sonnett, \& Wright}]{nugent2016neowise}
Nugent, C., Mainzer, A., Bauer, J., {et~al.} 2016, The Astronomical Journal, 152, 63

\bibitem[{Ostro {et~al.}(2006)Ostro, Margot, Benner, Giorgini, Scheeres, Fahnestock, Broschart, Bellerose, Nolan, Magri, {et~al.}}]{ostro2006radar}
Ostro, S.~J., Margot, J.-L., Benner, L.~A., {et~al.} 2006, Science, 314, 1276

\bibitem[{Pou \& Nimmo(2024)}]{pou2024tidal}
Pou, L., \& Nimmo, F. 2024, Icarus, 411, 115919

\bibitem[{Pravec {et~al.}(2024)Pravec, Meyer, Scheirich, Scheeres, Benson, \& Agrusa}]{pravec2024rotational}
Pravec, P., Meyer, A., Scheirich, P., {et~al.} 2024, Icarus, 116138

\bibitem[{Pravec {et~al.}(2023)Pravec, Scheirich, Kusnirak, Hornoch, Galád, \& Velen}]{pravecdatabase}
Pravec, P., Scheirich, P., Kusnirak, P., {et~al.} 2023, https://www.asu.cas.cz/~ppravec/neo.htm, Retrieved December 11, 2023

\bibitem[{Pravec {et~al.}(2006)Pravec, Scheirich, Ku{\v{s}}nir{\'a}k, {\v{S}}arounov{\'a}, Mottola, Hahn, Brown, Esquerdo, Kaiser, Krzeminski, {et~al.}}]{pravec2006photometric}
Pravec, P., Scheirich, P., Ku{\v{s}}nir{\'a}k, P., {et~al.} 2006, Icarus, 181, 63

\bibitem[{Pravec {et~al.}(2016)Pravec, Scheirich, Ku{\v{s}}nir{\'a}k, Hornoch, Gal{\'a}d, Naidu, Pray, Vil{\'a}gi, Gajdo{\v{s}}, Korno{\v{s}}, {et~al.}}]{pravec2016binary}
---. 2016, Icarus, 267, 267

\bibitem[{Pravec {et~al.}(2021)Pravec, Scheirich, Scheeres, McMahon, Meyer, Ku{\v{s}}nir{\'a}k, Hornoch, Ku{\v{c}}{\'a}kov{\'a}, Fatka, McMillan, {et~al.}}]{pravec2021photometric}
Pravec, P., Scheirich, P., Scheeres, D., {et~al.} 2021, in 7th IAA Planetary Defense Conference, 24

\bibitem[{Quillen {et~al.}(2022)Quillen, LaBarca, \& Chen}]{quillen2022non}
Quillen, A.~C., LaBarca, A., \& Chen, Y. 2022, Icarus, 374, 114826

\bibitem[{Rubincam(2000)}]{rubincam2000radiative}
Rubincam, D.~P. 2000, Icarus, 148, 2

\bibitem[{Scheeres(2006)}]{scheeres2006relative}
Scheeres, D.~J. 2006, Celestial Mechanics and Dynamical Astronomy, 94, 317

\bibitem[{Scheeres(2009)}]{scheeres2009stability}
---. 2009, Celestial Mechanics and Dynamical Astronomy, 104, 103

\bibitem[{Scheeres(2016)}]{scheeres2016orbital}
---. 2016, Orbital motion in strongly perturbed environments: applications to asteroid, comet and planetary satellite orbiters (Springer)

\bibitem[{Scheeres {et~al.}(2021)Scheeres, McMahon, Bierhaus, Wood, Benner, Hartzell, Hayne, Jedicke, Le~Corre, Meyer, {et~al.}}]{scheeres2021janus}
Scheeres, D.~J., McMahon, J., Bierhaus, E.~B., {et~al.} 2021, in 7th IAA Planetary Defense Conference, 55

\bibitem[{Taylor \& Margot(2011)}]{taylor2011binary}
Taylor, P.~A., \& Margot, J.-L. 2011, Icarus, 212, 661

\bibitem[{Taylor \& Margot(2014)}]{taylor2014tidal}
---. 2014, Icarus, 229, 418

\bibitem[{Vokrouhlick{\`y} {et~al.}(2007)Vokrouhlick{\`y}, Breiter, Nesvorn{\`y}, \& Bottke}]{vokrouhlicky2007generalized}
Vokrouhlick{\`y}, D., Breiter, S., Nesvorn{\`y}, D., \& Bottke, W. 2007, Icarus, 191, 636

\bibitem[{Walsh {et~al.}(2008)Walsh, Richardson, \& Michel}]{walsh2008rotational}
Walsh, K.~J., Richardson, D.~C., \& Michel, P. 2008, Nature, 454, 188

\bibitem[{Wang \& Hou(2021)}]{wang2021break}
Wang, H.-S., \& Hou, X.-Y. 2021, Monthly Notices of the Royal Astronomical Society, 505, 6037

\bibitem[{Wisdom {et~al.}(1984)Wisdom, Peale, \& Mignard}]{wisdom1984chaotic}
Wisdom, J., Peale, S.~J., \& Mignard, F. 1984, Icarus, 58, 137

\end{thebibliography}
\bibliographystyle{aasjournal}



\end{document}